\documentclass[journal,dvipsnames,svgnames]{vgtc}                     % final (journal style)
\onlineid{1795}

\vgtccategory{Research}

\vgtcpapertype{application/design study}

\title{\systemname: \revision{Visualization of AI-Assisted Task Guidance in AR}}

\author{%
\authororcid{Sonia Castelo*}{0000-0001-6881-3006}, 
\authororcid{Joao Rulff*}{0000-0003-3341-7059}, 
\authororcid{Erin McGowan*}{0000-0002-7565-3052}, 
\authororcid{Bea Steers}{0009-0007-2831-6460}, 
\authororcid{Guande Wu}{0000-0002-9244-173X}, 
\authororcid{Shaoyu Chen}{0000-0002-1856-6294}, \\
\authororcid{Iran Roman}{0000-0003-3781-7244}, 
\authororcid{Roque Lopez}{0000-0003-3484-1783}, 
\authororcid{Ethan Brewer}{0000-0002-4367-0763},
\authororcid{Chen Zhao}{0000-0001-7672-146X}, 
\authororcid{Jing Qian}{0000-0002-5517-3035},  \\
\authororcid{Kyunghyun Cho}{0000-0003-1669-3211}, 
\authororcid{He He}{0009-0005-7805-3780},  
\authororcid{Qi Sun}{0000-0002-3094-5844}, 
\authororcid{Huy Vo}{0000-0002-5963-6615}, 
\authororcid{Juan Bello}{0000-0001-8561-5204}, 
\authororcid{Michael Krone}{0000-0002-1445-7568}, and 
\authororcid{Claudio Silva}{0000-0003-2452-2295}
}
\authorfooter{All authors are with the New York University. E-mails: \{s.castelo, jlrulff, erin.mcgowan, bs3639, guandewu, sc6439, irr2020, rlopez, ethan.brewer, cz1285, jq2267, kyunghyun.cho, hh2291, qs2053, huy.vo, jpbello, mk8949, csilva\}@nyu.edu. * These authors contributed equally and should be regarded as joint first authors.}

\abstract{%
    The concept of augmented reality (AR) assistants has captured the human imagination for decades, becoming a staple of modern science fiction. 
    \revision{To pursue} this goal, it is necessary to develop artificial intelligence (AI)-based methods that simultaneously perceive the 3D environment, reason about physical tasks, and model the \revision{performer}, all in real-time. 
    Within this framework, a wide variety of sensors are needed to generate data across different modalities, such as audio, video, depth, speech, and time-of-flight. 
    The required sensors are typically part of the AR headset, providing \revision{performer} sensing and interaction through visual, audio, and haptic feedback. 
    AI assistants not only record the \revision{performer} as they perform activities, but also require machine learning (ML) models to understand and assist the \revision{performer} as they interact with the physical world. 
    Therefore, developing such assistants is a challenging task. 
    \revision{We propose ARGUS,} a visual analytics system to support the development of intelligent AR assistants.     
    Our system was designed as part of a multi-year-long collaboration between visualization researchers and ML and AR experts. 
    This co-design process has led to advances in the visualization of ML in AR. 
    \revision{Our system allows for online visualization of object, action, and step detection} as well as offline analysis of previously recorded AR sessions. % containing video, audio, depth map, and inertial measurement unit (IMU) data. 
    \revision{It visualizes} not only the multimodal sensor data streams but also the output of the ML models. 
    This allows developers to gain insights into the \revision{performer activities} as well as the ML models, helping them troubleshoot, improve, and fine-tune the components of the AR assistant.
}


\teaser{
  \centering
  \includegraphics[width=0.9\linewidth]{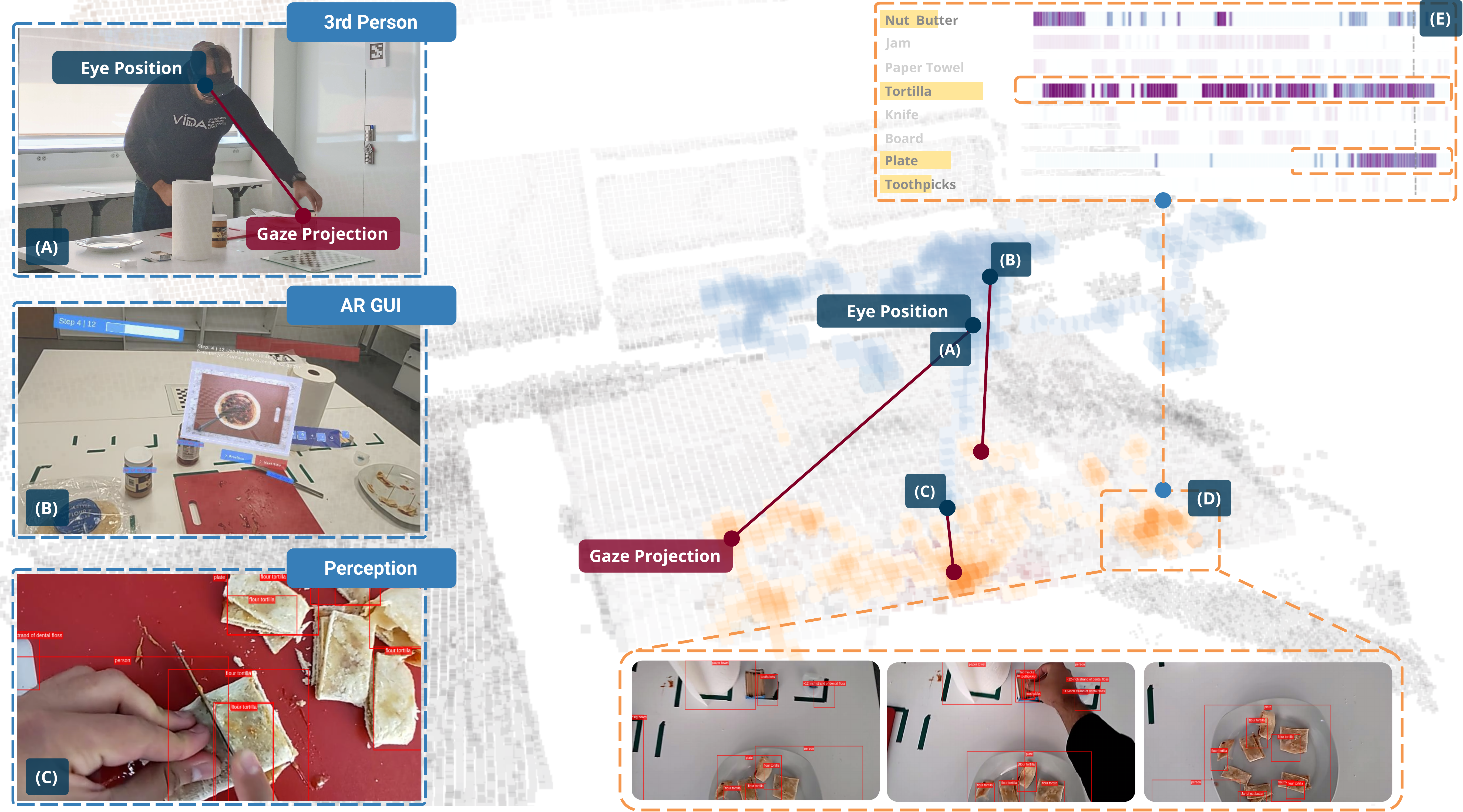}
  \caption{\revision{\systemname is a visual analytics tool for real-time and historical evaluation of sensor and model outputs of AR task assistants. Shown here are (A) 3rd-person perspective of a human (\emph{performer}) performing a task with AR headset guidance, (B) the AR GUI from the perspective of the performer, (C) a snapshot of a visual perception model analyzing data from the headset camera. For each time step, detailed information highlights the performer's gaze direction and the corresponding frame recorded by the headset. (D) heatmaps of the performer's gaze projection onto the world point cloud allow for inspection and understanding of their attention over time. The cluster on the left shows the performer finalizing a recipe. (E) object detection information from \systemname Temporal View illustrating that \emph{Plate} was only detected towards the end of the task while \emph{Tortilla} was detected throughout the whole session.}
  }
  \label{fig:teaser}
}

\graphicspath{{figs/}{figures/}{pictures/}{images/}{./}} % where to search for the images

\newcommand{\systemname}{\emph{ARGUS}\xspace}
\newcommand{\systemnamelong}{Augmented Reality Guidance and User-modeling System\xspace}

\usepackage{tabu}                      % only used for the table example
\usepackage{booktabs}                  % only used for the table example
\usepackage{lipsum}                    % used to generate placeholder text
\usepackage{mwe}                       % used to generate placeholder figures
\usepackage{color}
\usepackage{soul}
\usepackage{ifthen}
\usepackage{mathptmx}                  % use matching math font
\usepackage{comment}

\newcommand{\critical}[1]{\tikz[baseline=(X.base)]\node [draw=red,fill=red!30,semithick,rectangle,inner sep=2pt, rounded corners=3pt] (X) {Critical};}

\newcommand{\important}[1]{\tikz[baseline=(X.base)]\node [draw=yellow,fill=yellow!30,semithick,rectangle,inner sep=2pt, rounded corners=3pt] (X) {Important};}

\newcommand{\optional}[1]{\tikz[baseline=(X.base)]\node [draw=green,fill=green!30,semithick,rectangle,inner sep=2pt, rounded corners=3pt] (X) {Optional};}

\newcommand\myparagraph[1]{ \vspace{4pt} \noindent \textbf{#1.}}
\newcommand\myparagraphit[1]{ \vspace{4pt} \noindent \textit{#1.}}

\newcommand{\final}{0}
\newcommand\revision[1]{{#1}}
\newcommand{\highlight}[1]{{\color{red} [{#1}]}}
\newcommand{\sonia}[1]{{\color{red} Sonia: [{#1}]}}
\newcommand{\erin}[1]{{\color{blue} Erin: [{#1}]}}
\newcommand{\joao}[1]{{\color{purple} Joao: [{#1}]}}
\newcommand{\ethan}[1]{{\color{brown} Ethan: [{#1}]}}
\newcommand{\michael}[1]{{\color{orange} Michael: [{#1}]}}
\newcommand{\shaoyu}[1]{{\color{pink} Shaoyu: [{#1}]}}
\newcommand{\jing}[1]{{\color{teal} Jing: [{#1}]}}
\newcommand{\iran}[1]{{\color{cyan} Iran: [{#1}]}}
\newcommand{\bea}[1]{{\color{cyan} Bea: [{#1}]}}
\newcommand{\qi}[1]{{\color{olive} Qi: [{#1}]}}
\newcommand{\chen}[1]{{\color{magenta} Chen: [{#1}]}}
\newcommand{\guande}[1]{{\color{violet} Guande: [{#1}]}}
\newcommand{\roque}[1]{{\color{blue} Roque: [{#1}]}}
\newcommand{\claudio}[1]{{\color{red} Claudio: [{#1}]}}
\newcommand{\warning}[1]{{\color{red} Warning: [{#1}]}}

\ifthenelse{\equal{\final}{1}}
{
\renewcommand{\highlight}[1]{}
\renewcommand{\sonia}[1]{}
\renewcommand{\erin}[1]{}
\renewcommand{\joao}[1]{}
\renewcommand{\ethan}[1]{}
\renewcommand{\michael}[1]{}
\renewcommand{\shaoyu}[1]{}
\renewcommand{\jing}[1]{}
\renewcommand{\iran}[1]{}
\renewcommand{\bea}[1]{}
\renewcommand{\qi}[1]{}
\renewcommand{\chen}[1]{}
\renewcommand{\guande}[1]{}
\renewcommand{\roque}[1]{}
\renewcommand{\claudio}[1]{}
\renewcommand{\warning}[1]{}
}
{}

\begin{document}

%%%%%%%%%%%%%%%%%%%%%%%%%%%%%%%%%%%%%%%%%%%%%%%%%%%%%%%%%%%%%%%%
%%%%%%%%%%%%%%%%%%%%%% UNCOMMENT if need urgent space saving %%%%%%%%%%%%%%%%%%%%%%
%%%%%%%%%%%%%%%%%%%%%%%%%%%%%%%%%%%%%%%%%%%%%%%%%%%%%%%%%%%%%%%%
%\setlength{\titlespace}{0ex}
%\setlength{\titlespace}{1.0ex}
%\setlength{\textheight}{9.25in}
\setlength{\abovecaptionskip}{1.0ex}
\setlength{\belowcaptionskip}{1.0ex}
\setlength{\floatsep}{1.0ex}
\setlength{\dblfloatsep}{\floatsep}
\setlength{\textfloatsep}{2.0ex}
\setlength{\dbltextfloatsep}{\textfloatsep}
\setlength{\abovedisplayskip}{1.0ex}
\setlength{\belowdisplayskip}{1.0ex}
%\setlength{\arraycolsep}{1ex}
%\setlength{\parskip}{1.0ex} % actually increase space for acmart
%\setlength{\baselineskip}{1ex}
%\setlength{\teaserspace}{1ex}
%\linespread{1.0} % default 1.0; WMD! very evil, be very careful

%%%%%%%%%%%%%%%%%%%%%%%%%%%%%%%%%%%%%%%%%%%%%%%%%%%%%%%%%%%%%%%%
%%%%%%%%%%%%%%%%%%%%%% START OF THE PAPER %%%%%%%%%%%%%%%%%%%%%%
%%%%%%%%%%%%%%%%%%%%%%%%%%%%%%%%%%%%%%%%%%%%%%%%%%%%%%%%%%%%%%%%

\firstsection{Introduction}
\maketitle
\label{sec:introduction}

% \joao{Performers: people using Hololens for tasks. Users: people using \systemname to explore the data}

The concept of an augmented reality (AR) assistant has captured the human imagination for years, becoming a staple of modern science fiction through popular franchises such as the \textit{Marvel Cinematic Universe}, \textit{Star Trek}, and \textit{Terminator}. 
The applications of such a system are seemingly endless. Humans, even those with domain expertise, are fallible creatures with imperfect memories whose skills deteriorate over time, especially during repetitive tasks or under stress. 
An AR assistant could help experts and novices alike in performing both familiar and new tasks. For instance, an AR assistant could aid a surgeon performing a familiar yet complex procedure, who could benefit from a second set of ``eyes'' due to the high-stakes nature of their task. 
Equally, it could walk an amateur chef through the steps of an unfamiliar recipe. In an ideal scenario, the AR assistant would become ``invisible'' in the sense that it is seamlessly integrated into the task procedure, providing well-timed audio and visual feedback to guide uncertain \revision{performers} and correct human errors while otherwise fading into the background. 
\revision{Overall}, the AR assistant would be able to reduce human error via correction, improve performance by reducing cognitive load, and introduce new tasks across a wide variety of applications. 

While \revision{aspects} of this vision \revision{are} currently \revision{still} aspirational, we are finally beginning to develop the technology that allows concepts once relegated to the world of science fiction to become reality. 
With respect to machine perception, the recent explosion of research on machine learning (ML), especially deep neural networks, has given way to powerful models able to detect objects, actions, and speech in real %removed revision hyphen here since "real time" is being used as a noun (and not an adjective/hyphenated modifier)
time with high accuracy. 
Ever-evolving implementations of Bayesian neural networks, reinforcement learning, and dialog systems (\revision{e.g.,} conversational agents) allow for task modeling and transactional question answering. 
A rise in AR technology, especially the commercial availability of headsets such as Microsoft HoloLens~2, Magic Leap, Google Glass, or Meta Quest Pro \revision{(and soon, Apple Vision Pro)} \revision{has provided} the hardware necessary for task guidance. 
The time is ripe for the development of assistive AR systems.\looseness=-1

\myparagraph{Challenges in perceptually-enabled task guidance} 
Developing an AR assistant, however, comes with a host of challenges. Such a system requires several moving parts to work in tandem to perceive the performer's environment and actions, reason through the consequences of a given action, and interact with both the performer and the user (for the sake of clarity, we will refer to subjects using the AR system to perform tasks during a session as ``performers'' and subjects using \systemname to collect and analyze data as ``users''). Creating these parts is a complex, time and computational resource-consuming process. 
The challenges include collecting, storing, and accessing a large volume of annotated data for model training, real-time sensor data processing for action and object recognition (or reasoning), and \revision{performer behavior} modeling based on first-person perspective data collected by the AR headset (see Sec.~\ref{sec:requirements} for a more detailed discussion of tasks and requirements). 

% \michael{I moved the longer list describing the challenges to \autoref{sec:requirements} to make the intro shorter and easier to read. Would be good if someone could cross-check if this makes sense.} \joao{I agree. The intro reads better now} \michael{Clearly mention here that there are two modes of operation for \systemname: the online (user monitoring) and offline (visual analytics / analysis of historical data) mode.} \joao{Done. Please check.}

\myparagraph{Our Approach} We propose \textbf{\systemname: \systemnamelong}, a visual analytics tool that facilitates multimodal data collection, enables modeling of \revision{the physical state of the environment and performer behavior}, and allows for retrospective analysis and debugging of historical data generated by the AR sensors and ML models that support task guidance. Our tool operates in two main modes. The \hyperref[fig:onlinedebugger]{online mode} (\revision{see} Sec.~\ref{subsec:onlinemode}) supports real-time monitoring of \revision{model} behavior and data acquisition during task execution time. This mode displays tailored visuals of real-time model outputs, which allows users of \systemname to monitor the system during live sessions and facilitates online debugging. Data is saved incrementally. Once finalized, all data and associated metadata collected during the task is seamlessly \revision{stored to} permanent data store with analytical capabilities able to handle both structured data generated by ML models and multimedia data (e.g. video, depth, and audio streams) collected by the headset. 
% \joao{What are the preprocessing steps taking place right before moving the data to the historical DB. Should we highlight this process?}. 
Our system can be used to explore and analyze historical session data by interacting with visualizations that summarize spatiotemporal information as well as highlight detailed information regarding model performance, performer behavior, and the physical environment (see Sec.~\ref{subsec:offlinemode}).

Our design was inspired by requirements from developers of AR systems and experts that create and evaluate these systems in the context of \revision{the} Defense Advanced Research Projects Agency's (DARPA) Perceptually-enabled Task Guidance (PTG) program \cite{ptg_site}. These experts \revision{use} \systemname and \revision{have} provided feedback throughout its development. In summary, \textbf{our main contributions are:}

\begin{itemize}

\item \systemname, a visual analytics tool tailored to the development and debugging of intelligent assistive AR systems. It supports online monitoring during task execution as well as retrospective analysis and debugging of historical data by coupling a scalable data management framework with a novel multimodal visualization interface \revision{capable of uncovering interaction patterns between performer actions and model outputs.} 

\item The design of novel visual representations to support complex spatiotemporal analysis of heterogeneous, multi-resolution \revision{data (i.e., data streams with different frame rates)}. 
\systemname \revision{not only} supports the visualization of internal AR assistant ML states in the context of the actions of the performer, \revision{but also the visualization of the interactions of the performer with the physical environment.}

% \item A list of requirements elicited in partnership with assistive technology model developers that such a tool must satisfy.

\item \revision{We demonstrate the usefulness of \systemname by a} set of case studies that demonstrate real-world use of \systemname, exhibiting how AR assistant developers \revision{leverage} the tool to improve their systems. 
\end{itemize}

% \noindent

% \revision{\textbf{Paper structure.} 
\revision{This paper is organized as follows: Sec.~\ref{sec:related_work} reviews the relevant literature on assistive AR systems and visualization of related data. Sec.~\ref{sec:overview} provides background and context for \systemname, including the AR personal assistant framework and architecture it is designed upon. Sec.~\ref{sec:requirements} specifies the requirements we aim to achieve. Sec.~\ref{sec:system} describes \systemname in detail, including all components of its online real-time debugging mode and offline data analytics mode. Sec.~\ref{sec:case_studies} explores two case studies in which \systemname proves useful to AR task assistant developers, ending with user feedback and limitations of our system.  Finally, we offer concluding remarks and future work in Sec.~\ref{sec:conclusion}.}

\section{Related Work} 
\label{sec:related_work}

\subsection{Assistive AR Systems}
The idea of using AR technologies to build assistive systems that have an internal model of the real world and are able to augment what \revision{a} performer sees with virtual content dates back more than three decades~\cite{caudell_augmented_1992}.
\revision{Yet} only recent advances in AR display technologies and artificial intelligence \revision{(AI)}, combined with the processing power to run the necessary computations in real time, \revision{have} enabled us to start building such systems.
Referring to the terminology introduced by Milgram and Kishino~\cite{milgram_taxonomy_1994} in their seminal paper on Mixed Reality, this not only requires a \emph{class 3 display}---a head-mounted display (HMD) equipped with see-through capability that can overlay virtual content on top of the real world---but also a great \emph{extent of world knowledge}\revision{. That is, the environment} should be modeled as completely as possible so that the assistive system can react to objects and actions in the real world.
Simultaneously, the \emph{reproduction fidelity} and the \emph{extent of presence} of an assistive system should be minimal, since the performer needs to focus on the real world, not be immersed in virtual content. In addition, the in-situ instructions help to reduce errors and facilitate procedural tasks. 
To date, results are mixed for task completion time using an assistive AR system versus not, with several studies finding longer times with assistive AR systems~\cite{tang2003comparative,zheng2015eye} \revision{whereas} others find the opposite~\cite{funk2015using}. \revision{Nevertheless}, most studies agree that AR helps to reduce errors and overall cognitive load as it provides in-situ instruction and guidance.\looseness=-1

AR can be enabled by a multitude of different display technologies, ranging from handheld devices like smartphones \revision{and} tablets to projector-based solutions \revision{and} heads-up displays found in airplanes or modern cars.
\revision{We, however,} focus on see-through AR HMDs for assistive AR systems, since these \revision{do} not significantly \revision{encumber} the performer. \revision{These headset displays do not restrict performers} to a limited space and \revision{leave} their hands free to execute situated tasks in the real world.
Furthermore, they usually offer a wider range of built-in sensors for modeling the environment and performer such as cameras, microphones, or IMUs.
See-through AR headset displays available today \revision{include} Microsoft HoloLens~2 (the hardware platform used in our work) \revision{and} Magic Leap~2.

As \revision{was} proposed by Caudell and Mizell~\cite{caudell_augmented_1992}, a common use case for such systems is to support \revision{performers} in repair and maintenance tasks~\cite{henderson_exploring_2011, fernandez_del_amo_augmented_2018}.
Similarly, AR assistants were proposed for manufacturing, e.g., training~\cite{liu_smart_2018} or live monitoring of production lines~\cite{becher_situated_2022}.
Another prominent area for AR assistive systems is healthcare and medicine~\cite{beams_evaluation_2022}, e.g., to assist surgery~\cite{puladi_augmented_2022} or other procedures~\cite{sun_high-precision_2020, jiang_hololens-based_2020}.
Furthermore, digital assistants can also make use of AR to enable a virtual embodiment of the assistant~\cite{schmeil_mara_2007, nijholt_towards_2022, kim_does_2018}.
Most of the modern systems mentioned above integrate ML methods for specific tasks, e.g., for object or voice command recognition.
However, they are mostly tailored to specific tasks and only have limited support for situated \revision{performer} modeling and perceptual grounding.
Integrating more complex AI methods will make the development and testing of such systems also more challenging.

To support the development of AR assistants, software toolkits have been proposed, for example, RagRug~\cite{fleck_ragrug_2022}, which is designed for situated analysis, or \revision{Data} visualizations in eXtended Reality (DXR)~\cite{sicat_dxr_2019}, which is specifically designed to build immersive analytics~\cite{marriott_immersive_2018} applications.
However, while such toolkits make it easier to develop feature-rich assistive systems that use data from the multiple sensors provided by the AR headset display and integrate AI methods, they do not offer explicit tools for external debugging of the required ML models and sensor streams.
Our goal is to fill this gap with \systemname.
This requires visualizing the multiple data streams from the sensors as well as the output of the models.

\subsection{Visualization of Multivariate Temporal Data}
The visualization of multivariate temporal data is a very active field of research.
A plethora of different methods and tools have been proposed which, for example, use multiple views, aggregation, and level-of-detail visualizations to represent the data efficiently.
A review of these methods is beyond the scope of this paper, therefore, we refer to a number of comprehensive surveys~\cite{aigner_visualization_2011, kehrer_visualization_2013, liu_visualizing_2017}.

\begin{comment}
Some methods use unsupervised ML methods for visualizing multivariate temporal data. 
For example, Brich et al.~\cite{brich_visual_2022} presented a visual analytics application for sensor streams of patients in intensive care units.
They visualize the multivariate temporal data as TimeCurves~\cite{bach_time_2016}, which use dimensionality reduction to project each time step from the high-dimensional data into a low-dimensional space (usually 2D).
That is, distances between points in the embedding do not show temporal difference as on a timeline, but the similarity of the data values.
A similar approach was also used by Bernard et al.~\cite{bernard_visual-interactive_2019}.
Since the temporal order of events is an important feature for debugging, this approach is unfeasible for our intended application.

Another class of visual analytics methods for multivariate temporal data focuses on data mining and pattern detection.
A recent example is PSEUDo by Yu et al.~\cite{yu_pseudo_2023}, which uses locality-sensitive hashing to extract patterns in multidimensional data.
Since we are not interested searching for specific patterns, but want to enable an unbiased exploration of the data, such approaches are not feasible for \systemname.
\end{comment}

There have been recent attempts to develop visualization systems to debug and understand the data acquired by 
\revision{multimodal, integrative-AI applications}. 
PSI Studio, a platform to support the visualization of multimodal data streams~\cite{psi} is able to provide useful visualization of sets of recorded sessions. However, it requires the user to not only compose their own visual interfaces by organizing \revision{predefined} elements in a visualization canvas, but also to structure the streaming data in a predefined format, \textit{psi-store}.
Built with \revision{a similar} goal, Foxglove~\cite{foxglove_2023} requires developers to organize their data into a Robot Operating System (ROS) environment. 
Moreover, these tools \revision{focus on} supporting the visualization of the data streams and are not able to summarize long periods of recordings with visualizations. \revision{To the best of our knowledge, existing tools also lack the ability to debug associated ML models.}
Other visualization tools, such as Manifold~\cite{zhang_manifold_2019}, are tailored to the interpretation and debugging of general ML models. In our case, we are interested in a narrower set of ML models, those that pertain to the understanding the behavior of AI assistants, which have different requirements than other visualization systems.

%%%%%%%%%%%%%%%%%%%% COMMENTS %%%%%%%%%%%%%%%%%%%%

% \michael{\textbf{idea for Rel Work structure:}
% assistive AR systems now possible: modern hardware is quite capable (HoloLens etc.), literature on frameworks to build apps and actual applications (healthcare, repair and maintenance, etc.) -- use, e.g., object recognition etc., but currently often not very "intelligent", require better integration of AI methods (human-AI interaction, user modeling, etc.) -- development of more advanced intelligent AR assistants require new debugging tools (not included in current frameworks like RagRug or DXR) -- discuss current tools for multimodal data streams visualization (and what is missing).}

% \michael{\emph{Further potentially interesting areas of related work, which should maybe be mentioned (at least briefly, citing 1-2 works):} \textbf{Video Analytics:}\\
% video and eye tracking visualization, e.g, Kurzhals et al.~\cite{kurzhals_iseecube_2014, kurzhals_gaze_2016}, can be combined with sensemaking~\cite{tanisaro_visual_2015}\\
% automated, ML-based video summarization~\cite{kumar_f-_2018, kadam_recent_2022} (this could also be mentioned as future work, I guess) \textbf{human-AI interaction} \cite{wienrich_extended_2021} review/perspective paper that mentions AR, e.g., for embodiment }
\section{Background: Building The TIM Personal Assistant}
\label{sec:overview}

In this section we describe the context of the development of \systemname. 
This includes the ecosystem of components needed to support intelligent AR assistant systems, \revision{ranging} from software running on \revision{the} headset device to data management modules able to ingest data in real-time.

\subsection{Motivating Context}
The development of \systemname is driven in large part by the requirements of \revision{the} DARPA \revision{PTG} program \cite{ptg_site}. %both DARPA and PTG are defined in the intro section
PTG aims to develop AI technologies \revision{that} help users perform complex physical tasks while making them \revision{both} more versatile by expanding their skillset and more proficient by reducing their errors.
%\cite{ptg_site}
\revision{Specifically, the program} seeks to develop methods, techniques, and technology for \revision{AI} assistants that provide just-in-time visual and audio feedback to help with task execution. The goal is \revision{to} utilize wearable sensors (head-mounted cameras and microphones) that allow the AR assistant to see what the performer sees and hear what they hear, so that the assistant can provide helpful feedback to the performer through speech and aligned graphics. The assistants learn about tasks by ingesting knowledge from checklists, illustrated manuals, training videos, and other sources of information (e.g., making a meal from a recipe, applying a tourniquet from directions, conducting a preflight check from a checklist). They then combine this task knowledge with a perceptual model of the environment to support mixed-initiative and task-focused \revision{performer} dialogs. The dialogs may assist a \revision{performer} in completing a task, identify and correct error\revision{s} during a task, and instruct them through a new task, taking into consideration the \revision{performer}’s level of expertise.
As part of PTG, our team has been building TIM, the Transparent, Interpretable, and Multimodal AR Personal Assistant, which is described below.

\subsection{Overview of the TIM Personal Assistant}

Our assistive AR framework (TIM) integrates perceptual grounding, attention, and user modeling during real-time AR tasks, and is composed of multiple software and hardware components. TIM perceives the environment, including the state of the human performer, by using a variety of data streams (details below) which are the input to the task guidance system. TIM communicates with the performer through the HoloLens~2 headset display. 

The task guidance system is primarily composed of three AI components that interpret the incoming data streams:
% The perception component is responsible for characterizing and communicating the current state of the task and its entities.
% The memory component takes perception outputs and integrates them into a model of the 3D environment.
% Taking a look at the AI side, the dotted arrows on the right (see Figure~\ref{fig:architecture}) are to show how the traffic is actually flowing, but the solid arrows can just be followed to understand the data pipeline. 
(1) \emph{Percept\revision{ual Grounding}} utilizes information from historical instances of actions from similar tasks and makes its best prediction of what the current action and objects are. (2) \emph{\revision{Perceptual Attention}} takes the objects and transforms them into 3D coordinates and contextualizes objects over time in the 3D environment. (3) \emph{Reasoning} then uses the objects and actions returned by perception to identify which step of the task the user is in and to understand whether or not they are performing the task correctly. 
Any of these data can be ingested and displayed on our platform, \systemname. % , in the HoloLens AR user interface, or through command line tools.

% \textit{ARGUS} works with an assistive AR framework (TIM) that integrates perceptual grounding, attention and user modeling during real-time AR tasks. 

\subsection{System Architecture}
\label{subsec: dashboard_arquitecture}
% \bea{add here. please feel free to reorganize this section}

\begin{figure}
    \centering
    \includegraphics[width=\linewidth]{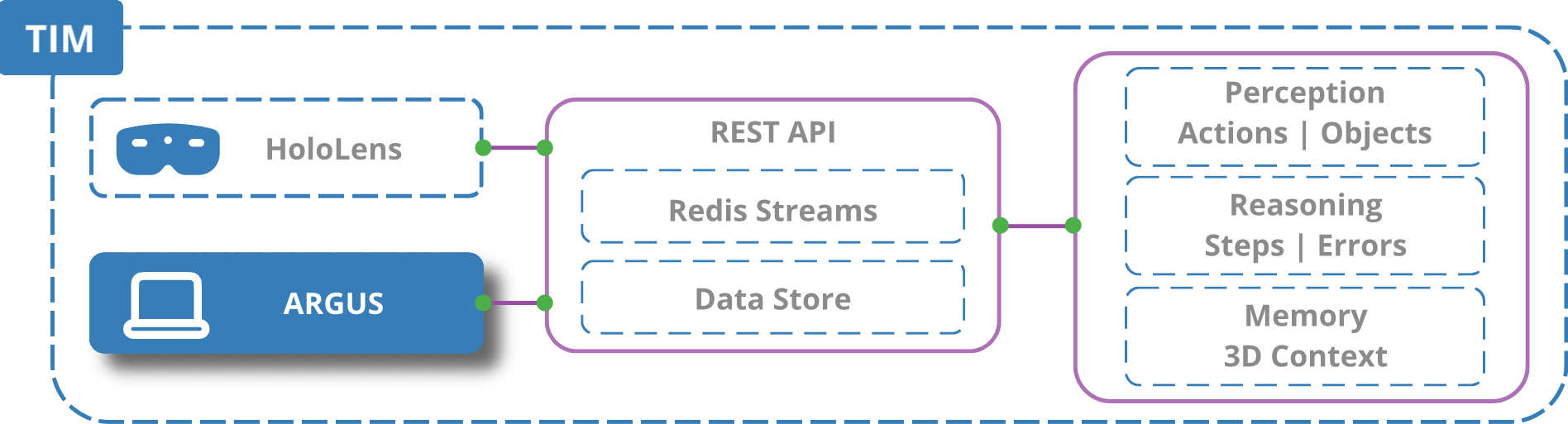}
    \vspace{-.2cm}
    \caption{\revision{TIM's architecture proposes a data communication service between the system components: the Hololens, AI modules, and \systemname.}
    }
    %\michael{Would be better readable with black text (and: can this be exported as a vector graphics (PDF)? Also, if we want to save some space, we could remove the TIM label and box.}
    \label{fig:architecture}
    \vspace{-.2cm}
\end{figure}

Since the computational resources on HoloLens~2 are limited, TIM is implemented as a client-server architecture. 
To enable data streaming capabilities, the system utilizes server-side infrastructure that provides a centralized data communication hub and real-time ML inference to facilitate ingesting, operating over, and contextualizing the produced data streams. A system diagram can be seen in Fig.~\ref{fig:architecture}.
% a message-hub for inter-component communication, as well as different computational modules \bea{bea fix wording} that provide machine learning and contextualize the data in the 3D environment.

% The high-level architecture of \systemname \michael{This seems to be confusing to me. \systemname is the debugging tool explained in \autoref{sec:system}, but we explain the AR assistant here, so should this say "The high-level architecture of TIM (Transparent, Interpretable, and Multimodal Personal Assistant)"? I guess it makes sense to explain the infrastructure here, but I would not mention \systemname already.} is depicted in Figure~\ref{fig:architecture}. 
% In what follows, we describe its key components.
%
%

% \myparagraph{Human Interfaces}
% \systemname is one of the human interfaces that allow the user to explore the data. Besides \systemname, HoloLens user and other developer tools are supported.
% \michael{See above -- I think we should mention \systemname here yet, but just explain the rest of the architecture that is then later accessed/used by \systemname. This section here should just explain TIM and the rest of the infrastructure.}

\myparagraph{Data Orchestration and Storage}
The core of our architecture is Redis Streams, which we use as our data message queue. 
A REST + Websocket API provides a uniform abstraction layer for components to interact with. 
% that we wrap around Redis so that our other components can be independent to server-side internals.
The HoloLens streams its sensor data to the API where it is made available to all other components in the system. 
The user is able to record data streaming sessions, which will listen and copy all data streams to disk.
% If the user is recording, the recorders will listen and produce a copy of the data to disk.
Later, users can selectively replay that data in the system as if the HoloLens was running\revision{,} for easy offline testing.

\myparagraph{Communication}
TIM uses \revision{the} REST API to stream onboard sensor data (i.e., gaze, hand tracking; see details in Section \ref{subsec: data}) in real-time. This allows us to shift the computation-heavy tasks to the server while keeping the essential tasks on the HoloLens to improve responsiveness. TIM also collects the ML prediction results from the server and updates the AR interaction and interface accordingly. The AR client running on the HoloLens ingests two streams to support contextual interaction: a \textit{perception} stream that recognizes objects in the scene and a \textit{reasoning} stream that recognizes \revision{performer} actions. On average, these streams take about 100~ms to complete one update cycle to the AR client.

\subsection{Data}
\label{subsec: data}

The HoloLens 2 can provide various data from multiple sensors. 
With \emph{Research Mode} enabled~\cite{DBLP:journals/corr/abs-2008-11239}, we stream data from the main RGB camera, 4 grayscale cameras, an infrared depth camera, and an IMU that contains an accelerometer, gyroscope, and magnetometer. 
\revision{Details of the camera data} can be found in Table~\ref{table:data}. 
Although it is theoretically possible to stream some of the data at higher resolution or frame rate, the need to run a user interface on the HoloLens creates a practical limit. Not only are the computational resources limited, but streaming extra data consumes more energy and may result in headset overheating.

The streamed frame rate in practice may be lower due to the packet drop during streaming.
Hand tracking and eye tracking data are also streamed.
The eye tracking data consists of 3D gaze origin positions and directions. 
The hand tracking data consists of 26 joint points for each hand. 
Each joint point contains a 3D position and a quaternion that indicates the orientation.
In our system, the per-frame point cloud which consists of RGB and depth frames can be integrated into a holistic 3D environment. 
Performer sessions can vary in size. 
\revision{F}or instance, the recording of a simple recipe (preparing pinwheels\revision{~\cite{mitll_tasks}}) usually takes $\sim$6~min and results in $\sim$600\,MB data without the point cloud data, but 3\,GB with the point cloud data. 

\myparagraph{\revision{Privacy and Ethical Considerations}} \revision{While AR provides incredible opportunities, performer privacy must be protected during data collection and utilization \cite{pase2012ethical}.
%
%there are ethical considerations to take into account in terms of how data is being collected and used by the system~\cite{pase2012ethical}. One of the hardest problems is related to \revision{performer} privacy. 
%
Our experiment protocol is approved by an Institutional Review Board (IRB). 
It ensures data is never directly linked to an individual identity, code numbers, rather than names or other identifying information, are used for video recordings in \systemname. 
Names or any other identifiable information are not collected and do not appear in any part of the system. 
Despite these efforts, it is theoretically possible to re-identify \revision{performers}, see, e.g., \cite{nair2023truth}, where it is shown that motion data can be used for identification. Another path to re-identification is the audio produced by the voice interactions.
}

\begin{table}[]
    \caption{Description of streamed data from HoloLens~2 visual sensors.}
    \centering
    \begin{tabular}{lllr}
    \toprule
    Sensor                     & Resolution       & Format      & Framerate \\
    \midrule
    RGB camera                 & 760 $\times$ 428 & RGB8        & 7.5\,fps           \\
    Grayscale camera               & 640 $\times$ 480 & Grayscale8  & 1\,fps             \\
    Infrared active brightness & 320 $\times$ 288 & Grayscale8  & 5\,fps             \\
    Infrared depth             & 320 $\times$ 288 & Depth16     & 5\,fps \\
    \bottomrule
    \end{tabular}
    \label{table:data}
\end{table}

\subsection{AI Task Guidance System}
\label{subsec:modeling}

\myparagraph{Percept\revision{ual} Grounding} To connect what the HoloLens sees and hears to task knowledge, the AR assistant needs to be equipped with models to recognize physical objects, actions, sounds, and contexts needed to complete a specific task. TIM uses multimodal machine-sensing models to detect human-object interactions in the environment. The output is real-time estimations with model confidence levels of three environmental elements: object categories, object localizations, and human action detections. We modulate object outputs via text instructions, allowing us to selectively detect objects and actions that are part of a particular procedure (e.g., recipe) and disregard everything else. To achieve this, our models generate and compare text and sensor %(e.g., RGB camera) 
representations. The models have the following main features:

\myparagraphit{Object detection and localization} We use ``Detector with image classes'' (\revision{Detic}) \cite{zhou2022detecting} to generate these estimations, since it is a model that produces RGB frames and free-form text descriptions of objects of interest (e.g., ``the blue cup'') with bounding-box and object mask estimations for the regions in the frame where the objects are detected. Its direct comparison of RGB and text modalities is enabled via Contrastive Language-Image Pretraining (CLIP) \cite{radford2021learning}.

\myparagraphit{Action recognition} TIM supports three action-recognition models: Omnivore \cite{girdhar2022omnivore}, SlowFast, \cite{xiao2020audiovisual,kazakos2021slow,feichtenhofer2019slowfast} and EgoVLP \cite{qinghong2022egocentric}. These models process video streams to output verb-noun tuples that describe actions. Each model has its benefits and limitations. While Omnivore is considered state-of-the-art for action recognition, it is a classification model with a fixed vocabulary. EgoVLP has a joint RGB-text representation that, similar to Detic and CLIP, allows for the detection of free-form text descriptions of actions. SlowFast integrates audio and RGB information, potentially allowing for the detection of actions outside the RGB field of view. Therefore, the optimal model to use is dependent on the deployment conditions. 

\myparagraph{Reasoning and Knowledge Transfer}
Reasoning and knowledge transfer first preprocess the input task description and create the corresponding objects and actions needed for each step\revision{~\cite{zhang2012automatically,lin2020recipe}}. In each frame, it takes the object and action outputs from the perceptual component, along with the processed input task description, and makes two decisions. First, the reasoning module performs 
\textit{error detection}, in which it attempts to determine if the \revision{performer} has made an error in the current frame based on how much the objects and actions detected through perception align with the preprogrammed knowledge of the step. Second, it performs \textit{step prediction}, in which the system predicts whether the current step is complete and should move to the next step. This decision is governed by a hidden Markov model (HMM)\revision{~\cite{baum1966statistical}}-like approach that primarily uses the probability of each action to appear in a given step of the task. These probabilities are calculated beforehand on a training dataset.
\begin{figure*}[tb]
    \centering
    \includegraphics[width=\linewidth]{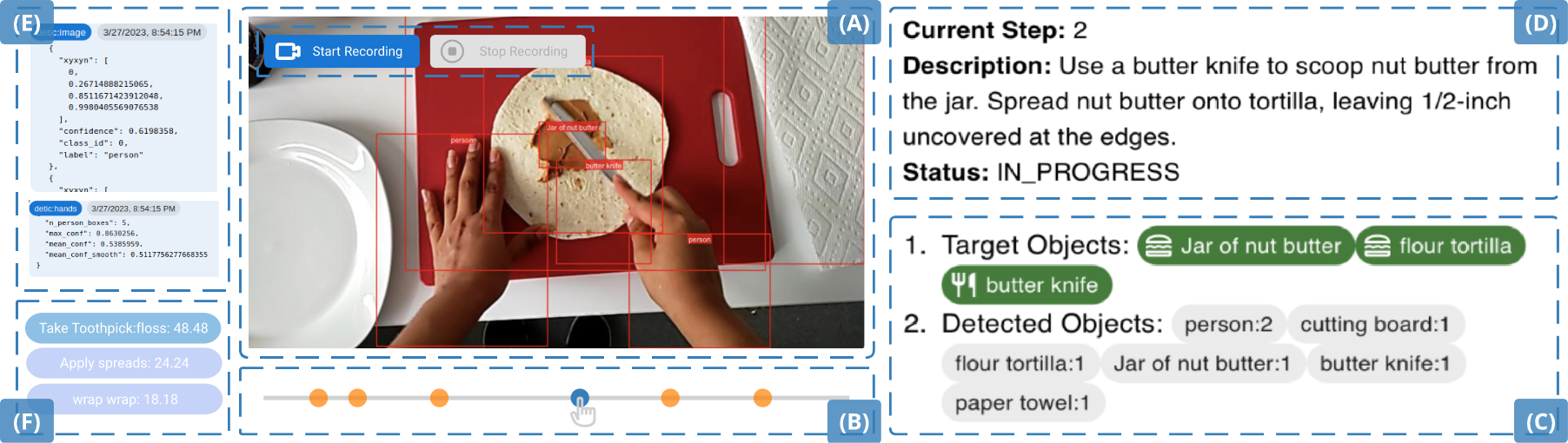}
    \caption{\revision{The online component of \systemname for real-time debugging. (A) Streaming Video Player: users can inspect the output of the headset’s camera overlayed with bounding boxes representing the detected objects. Users have the option to record any session. (B) Confidence Controller: a slider that allows the user to control the threshold model confidence. (C) Perception model outputs, including target and detected objects. ``Target Objects'' represent the objects needed in the current step (from recipe instructions) while ``Detected Objects'' shows all the objects identified by the perception models and their corresponding number of instances (e.g., multiple knives may be detected in a frame). (D) Reasoning model outputs, including the step and error predictions, step description, and the performer’s status. (E) Raw data views, showing the raw data collected by the system. (F) Widgets showing the predicted actions with their probabilities. In this example, the model predicts that the current action is ``Take Toothpick'' with 48\% likelihood, followed by ``Apply spreads'' with 24\% and "wrap wrap" with 18\%.}
    %\qi{I suggest explain A/B..E on.-by-one.}
    }
    \label{fig:onlinedebugger}
    \vspace{-.2cm} 
\end{figure*}
%%%%%%%%%%%%%%%%%%%%%%%%%%%%%%%%%%%%%%%%%%%%%%%%%%%%%%%%%%%%%%%%%%%%%%%%%%

\section{Tasks \& Requirements}
\label{sec:requirements}
\revision{\systemname was developed to support the development and operation of the AR personal assistant outlined in Sec.~\ref{sec:overview}.}
\revision{On top of the obvious need to visualize the multitude of raw data streams, \systemname was designed to enable the real-time and post-hoc \revision{visualization} of ML models and performer interactions in the context of the physical environment, all in a time-synchronized fashion.}
To summarize, such a system should have the following design requirements (R1-R5). These requirements were created by working side-by-side with the developers of the AR assistant components described in Section \ref{sec:overview} (i.e., perception, reasoning), and with end users 
% (i.e., pilots, cooks, medical technicians, surgeons) 
through interviews and feedback sessions during and after use of TIM and \systemname. For context, the AR-enabled tasks that \systemname aims to support include, but are not limited to: making a meal from a recipe, applying a tourniquet, repairing an engine, and completing an aircraft preflight check.

\begin{enumerate}[start=1,label={[\bfseries R\arabic*]}]

    %\important
    \item \label{req:live_monitoring} \textbf{Live monitoring}: The ability to visualize the output of the various components of the system during task execution. This is crucial to understand possible system failures before completing recording sessions and gaining real-time insights about model outputs.
    \vspace{-.15cm}
    %
    % \critical
    \item \label{req:provenance_storage} \textbf{Seamless provenance acquisition}: The future availability of the multimodal dataset produced during a recording session. This supports  developers in improving algorithms and debugging system outputs and researchers in retrospectively investigating user-generated data. Therefore, automatically storing the acquired data (and metadata) into databases % without requiring data wrangling efforts 
    is important for such a system.
    \vspace{-.15cm}
    % 
    %\important
    \item \label{req:model_debugging} \textbf{Retrospective analysis of model performance}:  The ability to visualize and inspect large chunks of the acquired data and model outputs to uncover \revision{relevant} spatial and temporal trends.
    \vspace{-.15cm}
    %
    %\important
    \item \label{req:physical_environment} \textbf{Physical environment representation}: A representation of the physical environment where the performance occurs. This representation should support data exploration tasks by explaining most of the observed user-generated data (e.g.\revision{, performer} movement patterns limited by physical constraints).
    \vspace{-.15cm}
    %
    %\critical
    \item \label{req:user_modeling} \textbf{Aggregated and detailed visualization \revision{performer} behavior}: A summary of the global interaction patterns of the user with the environment. This is key in analyzing general performer behavior. Aggregating large chunks of data temporally and spatially can hide important details, thus, the system should provide both global and local perspectives of \revision{performer} behavior data.  
\end{enumerate}

\section{\systemname: Augmented Reality Guidance and User-modeling System} 
\label{sec:system}

\begin{figure*}
    \centering
    \includegraphics[width=\linewidth]{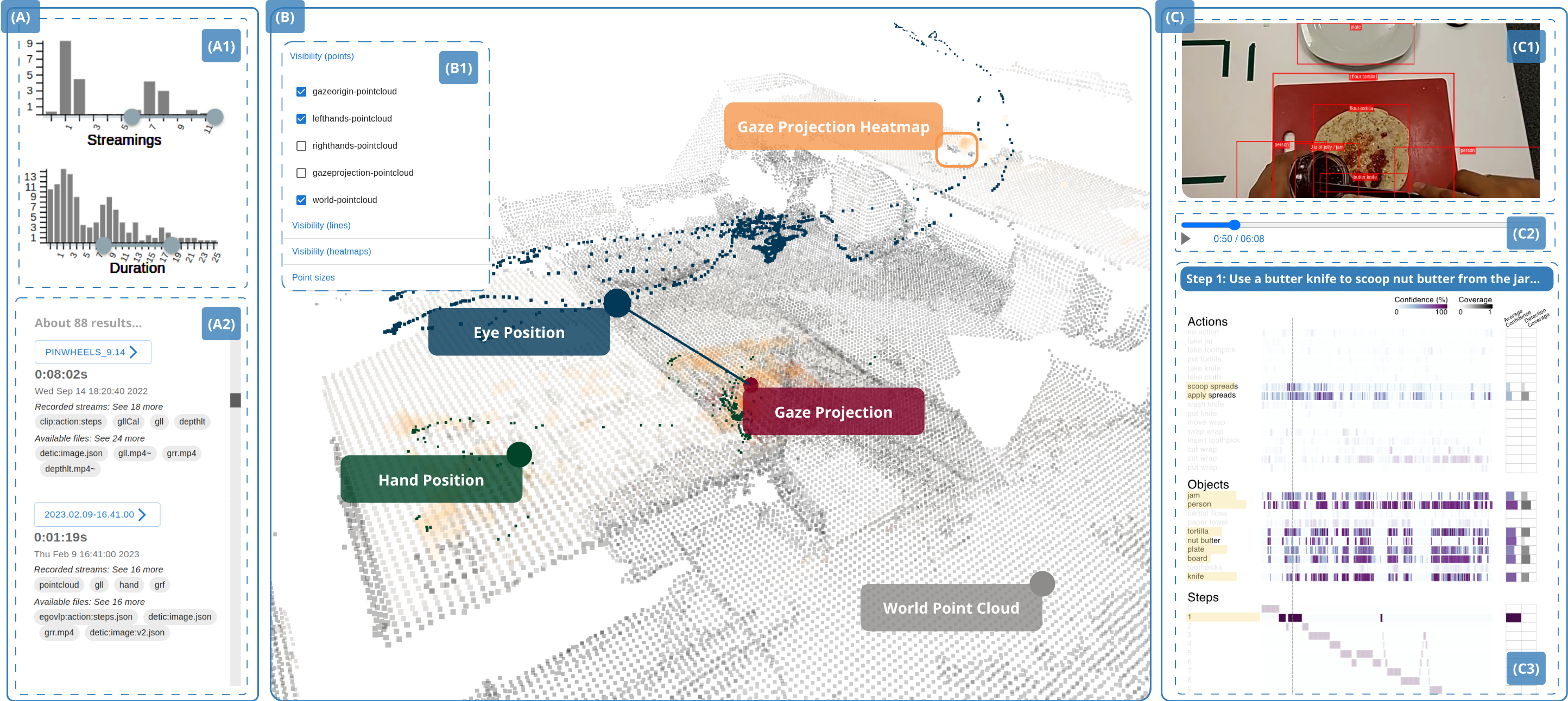}
    \caption{
    %Overview of the user interface of \systemname. (A) \textbf{Spatial View} shows the world point cloud representing the physical environment, 3D points for eye and hand positions, and gaze projections and heatmaps. (B) \textbf{Model Output Viewer} displays the output of the machine learning models (reasoning and perception) used during execution time. (C) \textbf{Video Frame} is the main camera video output of the current timestamp selected by the user. (D,E) \textbf{Data Manager} shows the applied filters and the list of retrieved sessions. (F) \textbf{Render Controls} allows the user to select the elements of the Spatial View they desire to see.
    \revision{Overview of the user interface and components of \systemname Offline. (A) The \textbf{Data Manager} shows the applied filters (A1) and the list of retrieved sessions (A2). 
    (B) The \textbf{Spatial View} shows the world point cloud representing the physical environment, 3D points for eye and hand positions, and gaze projections and heatmaps. (B1) Render Controls allow the user to select the elements of the Spatial View they desire to see. 
    (C) \textbf{Temporal View:}  (C1) The Video Player is the main camera video output of the current timestamp selected by the user. (C2) The Temporal Controller controls the video player and updates the model output viewer as well. (C3) The Model Output Viewer displays the output of the machine learning models (reasoning and perception) used during execution time. }
    }
    \label{fig:systemoverview}
\end{figure*}

As described in Sec.~\ref{sec:requirements}, we developed \systemname concomitantly with TIM to meet the development needs of building an effective AR task assistant. In total, \systemname enables the interactive exploration and debugging of all components of the data ecosystem needed to support intelligent task guidance. This ecosystem contains the data captured by the HoloLens's sensors and the outputs of the perception and reasoning models outlined in Sec.~\ref{sec:overview}. \systemname has two operation modes: ``Online'' (during task performance), and ``Offline'' (after performance). \revision{Users can use these two modes separately if needed, for instance, to perform real-time debugging through the \hyperref[fig:onlinedebugger]{online mode}. In another usage scenario, users may start by using the \hyperref[fig:onlinedebugger]{online mode} to record a session and then explore and analysis the data in detail using the \hyperref[fig:systemoverview]{offline mode}. We describe additional usage scenarios in Section \ref{sec:case_studies} through two case studies.}
\vspace{-0.05in}
\subsection{\systemname Online: Real-time Debugger}
\label{subsec:onlinemode}

\myparagraph{Real-Time Debugging} The \systemname architecture allows streaming data collection and processing in real-time, which makes instantaneous debugging and data validation possible \ref{req:live_monitoring}. As depicted in Fig.~\ref{fig:onlinedebugger}, the \hyperref[fig:onlinedebugger]{online mode} provides information on the outputs of the reasoning and perception models using custom visual widgets. The \revision{caption of} Fig.~\ref{fig:onlinedebugger} describes each component. Since what the HoloLens main camera sees (and thus what is analyzed by the models) is not the same as what the performer sees \revision{(due to different fields of view)}, having a real-time viewer such as (A) can help ensure the HoloLens is capturing what the performer and user wish to capture. Additionally, components \revision{(C) \& (F)} provide information that can help validate the objects and actions identified by the models in real-time (as opposed to having to do so post hoc). \revision{We note that these features are primarily intended to aid a user in analyzing performer behavior and model performance in real time, rather than to aid the performer as they complete a task.}  

\myparagraph{Data collection} Users of \systemname can decide when to save the recording for future analysis. By clicking the \textit{Start Recording} button, all data captured and generated by the sensors and models from that point on are redirected from the online streaming database to the historical database until the user clicks \textit{Stop Recording}. The data migration process is transparent to the user \ref{req:provenance_storage}.

%%%%%%%%%%%%%%%%%%%%%%%%%%%%%%%%%%%%%%%%%%%%%%%%%%%%%%%%%%%%%%%%%%%%%%%%%%
%%%%%%%%%%%%%%%%%%%%%%%%%%%%%%%%%%%%%%%%%%%%%%%%%%%%%%%%%%%%%%%%%%%%%%%%%%

\vspace{-0.05in}

\subsection{\systemname Offline: Visualizing Historical Data}
\label{subsec:offlinemode}

The \hyperref[fig:systemoverview]{offline mode's} main goal is to 
\revision{enable analysis} of historical data generated by the models \revision{and performer actions in the physical environment} \ref{req:model_debugging}. To allow for easy exploration of this large and heterogeneous data, \systemname provides a visual user interface that enables querying, filtering, and exploration of the data. Due to the spatiotemporal characteristics of the data, we provide both spatial and temporal visualization widgets to allow users to analyze the data from different perspectives. Fig.~\ref{fig:systemoverview} shows \revision{the} components composing \systemname in \hyperref[fig:systemoverview]{offline mode}.
In the following, we describe the main components of the \hyperref[fig:systemoverview]{offline mode}: the \hyperref[fig:systemoverview]{Data Manager}, the \hyperref[fig:systemoverview]{Temporal View}, and the \hyperref[fig:systemoverview]{Spatial View}. We highlight the interaction flow a user is \revision{likely} to follow, and for each component, we describe the visualizations, the interactions provided, and their goals.

\subsubsection{Data Manager}
\label{subsubsec:data_manager}

Users start the exploration by using the \hyperref[fig:systemoverview]{Data Manager} \revision{shown in} Fig.~\ref{fig:systemoverview}\revision{(A)} to filter the set of sessions available in the data store. Our data is organized as sessions (each session contains all recordings, data streams, and model outputs for a \revision{performer} executing a task). The \hyperref[fig:systemoverview]{Data Manager} enables data retrieval by allowing users to specify filters and select specific sessions from a list of results.

\myparagraph{Data Querying} Users can query the data by specifying various filters, as shown in Fig.~\ref{fig:systemoverview}\revision{(A1)}. Filters are presented in the form of histograms the users can brush to select the desired range.

\myparagraph{Query Results} The results component displays the retrieved sessions in a list format. Fig.~\ref{fig:systemoverview}\revision{(A2)} shows the results for a given query specified by the user. Each element represents a session showing key features, including name, duration, date, recorded streams, and available model outputs. Once an element of the list is \revision{selected}, the corresponding data will be loaded into the views of the system.

\subsubsection{Spatial View}
\label{subsubsec:spatial_view}

As described in Section \ref{subsec: data}, the spatial nature of some of the streamed data demands a 3D visualization to allow users to meaningfully explore the data. 
For this, \systemname provides a \hyperref[fig:systemoverview]{Spatial View} shown in Fig.~\ref{fig:systemoverview}\revision{(B)} that allows users to analyze how performers interact with the physical environment in conjunction with the spatial distribution of model outputs. The \hyperref[fig:systemoverview]{Spatial View} can help resolve where performers were located, where they were looking during specific task steps, where objects were located in the scene, etc.
Below, we describe the elements of the \hyperref[fig:systemoverview]{Spatial View} and its interaction mechanisms tailored to support the analysis of the spatial data following well-established visualization guidelines~\cite{shneiderman1996eyes} to provide both overview and detailed information.

The basis of the \hyperref[fig:systemoverview]{Spatial View} is a 3D point cloud (or \emph{world point cloud}) as shown in Fig.~\ref{fig:systemoverview}\revision{(B)} representing the physical environment where the performer is \revision{operating} %immersed 
\ref{req:physical_environment}. This representation helps us interpret different aspects of the space, such as physical constraints imposed by the environmental layout.
\revision{However,} the point clouds generated based on the data acquired by the headset cameras can easily contain millions of points, making it unfeasible to transfer them over the web and render them within most web browsers. 
%
%The point cloud generation process that takes place in Hololens~2 introduces redundancy in the final result, given that it generates approximately one point cloud per half second.
To give an overview of the whole task, all point clouds of one recording \revision{are} merged to obtain a temporal aggregation.
Hololens~2 generates approximately one point cloud per half second, which creates redundancy.
%based on the output of the RGB and depth cameras. 
This redundancy can be removed by creating a union of all point clouds and then downsampling it using voxelization. 
However, selecting imprecise parameters can lead to a subrepresentation of the physical space, losing important information and, consequently, hindering analysis.
%
%With this in mind
\revision{Thus}, we \revision{parameterize} the voxel-based downsampling to create voxels at 1\,\revision{cm} resolution, providing enough detail for the purposes of our tool. In our experiments, the downsampled point clouds had less than 100,000 points even in the worst case, leading to reasonable transfer and rendering times. With the world point cloud representing the physical environment, \revision{we are able to visualize} \revision{performer activity in context}. %\revision{Performance details are provided in Section~\ref{subsubsec:implementation}}.

As illustrated in Fig.~\ref{fig:systemoverview}, eye position, hand position, and other data streams can be represented as 3D points in the same scene. The blue dots show the eye position of the \revision{performer} during a session, while the green dots show the hand position. For each collection of 3D points representing a data stream, users can retrieve more detailed information by interacting with the points. 
For example, if the user hovers their mouse over the points representing the eye position, a line representing the gaze direction will automatically be rendered in the scene, representing what point in space the \revision{performer} was looking at from their current position at a specific timestamp. This is possible by calculating the intersection of the gaze direction vector with the world point cloud. This gaze information can also be represented as a 3D point cloud to provide a visual summary of the areas the performer was focused on \ref{req:user_modeling}. This interaction also updates the corresponding video frame in Fig.~\ref{fig:systemoverview}\revision{(C1)} and highlights the models' outputs in Fig.~\ref{fig:systemoverview}\revision{(C3)}.

Although the point cloud provides a summarization of the spatial distribution of these data streams, this representation fails to \revision{convey} aggregated statistics of the data, such as the density of points in a given region which is proportional to the amount of time the \revision{performer} spent in a given location of the scene. For this purpose, %a more suitable visualization, like a 3D heatmap, should be used.
\revision{a 3D heatmap is a more suitable visualization.}
The \hyperref[fig:systemoverview]{Spatial View} can create 3D heatmaps of each data stream. 
%Fig.~\ref{fig:systemoverview}A shows a heatmap with the distribution of the gaze data during the session. 
\revision{The heatmap in Fig.~\ref{fig:systemoverview}(B) shows the distribution of the gaze data during the session.}
We leverage the voxel information created during the downsampling %process 
to calculate the density of points within voxels. Using an appropriate color scale, we render cells with non-negligible densities to create the 3D heatmap. Every data stream containing spatial information can be incorporated into the \hyperref[fig:systemoverview]{Spatial View} as 3D point clouds or heatmaps in \systemname.
Information regarding perception and reasoning models are also available in this view. By combining bounding boxes generated by perception models and depth information captured by the headset, we reconstruct the center point of each \revision{detected object}, helping users understand the spatial distribution of objects. 
Also, occupancy maps representing the density of objects in different regions %of the space 
can be derived as presented in Fig.~\ref{fig:modelheatmap}.
Moreover, the \hyperref[fig:systemoverview]{Spatial View} provides a summarization of gaze information by rendering sets of vectors representing gaze directions over time. Users can control the style (e.g., size and opacity of points) and visibility of all data streams, choosing what data should be visible for analysis. Lastly, point clouds can also be filtered based on timestamp ranges, allowing for focused analysis of specific task steps (``Visibility'' in Fig.~\ref{fig:systemoverview}\revision{(B1)}).

\subsubsection{Temporal View}
\label{subsubsec:temporal_view}

ML models are a core component of an AI assistant system. While the field of ML has seen many recent advancements to support assistive AR applications~\cite{becher_situated_2022, puladi_augmented_2022, fernandez_del_amo_augmented_2018}, the need for tools to improve them remains. Model debuggers are powerful tools used to analyze, understand, and improve these models by identifying issues and probing ML response functions and decision boundaries. This helps developers mak\revision{e} models more accurate, fair, and secure, promoting trust and enabling understanding which is highly desirable in intelligent AR assistants. \systemname provides a model debugger based on temporal visualizations to debug the ML models used in AI assistant systems~\ref{req:user_modeling}. We describe the different temporal components in detail in the following subsections.

\begin{figure}[tbp]
    \centering
    \includegraphics[width=\linewidth]{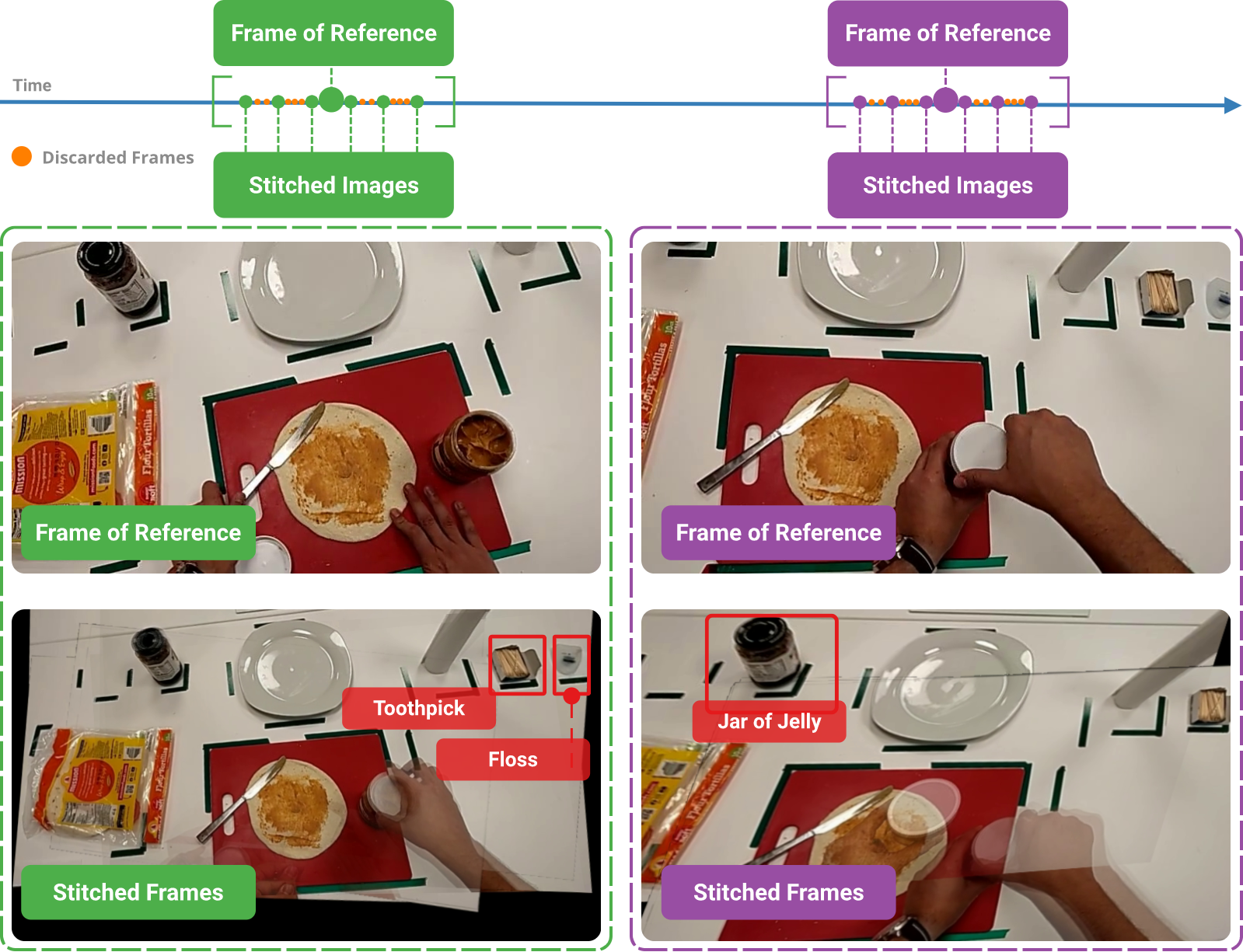}
    % \vspace{-.2cm} 
    \caption{\revision{A visual representation of frame selection for the panoramic mosaic view of the video player (top) and comparison of these panoramic mosaics with corresponding frames from the same timestep of the raw main camera video for reference (bottom). Each panoramic mosaic is composed of several frames sampled from a window around the current timestep of the raw main camera video. In both examples, we highlight objects that are visible in the panoramic mosaic but not in the raw main camera video (toothpick, floss, and jar of jelly, respectively) in red.}
    }
    \label{fig:stitching}
    % \vspace{-.1cm}
\end{figure}

\myparagraph{Video Player} The object detection model not only recognizes all objects in an image but also their positions. To inspect these outputs, \systemname contains a video player component that identifies the spatial location of detected objects over time, as shown in Fig.~\ref{fig:systemoverview}C. This component allows the user to toggle between two views: 1) the raw main camera video stream and 2) a \hyperref[fig:stitching]{panoramic mosaic view} which consists of a sequence of panoramic mosaics generated from this main camera stream. 
We highlight all detected objects with bounding boxes, which are provided by the object detection model.

%
%%%%%%%%%%%%%%%%%%%%%%%%%%%%%%%%%%%%%%%%%%%%%%%%%%%%%%%%%%%%%%%%%%%%%%%%%%
%%%%%%%%%%%%%%%%%%%%%%%%%%%%%%%%%%%%%%%%%%%%%%%%%%%%%%%%%%%%%%%%%%%%%%%%%%
%

%A typical method of visualizing the output of an object detection model is to simply draw bounding boxes over each individual frame of a video. \systemname provides such a visualization in the first view of the video player component.
%As this view displays the raw main camera video stream collected by the HoloLens, it 
The first view of the video player, which displays the raw main camera video stream collected by HoloLens,
enables a highly granular level of model debugging.
This allows the user to note specific frames where object detection failed or yielded unexpected results. 
However, the main camera of the HoloLens has a limited field of view. Often objects that the performer sees at a given timestep cannot be seen in the frame of the raw main camera video at that timestep. 
Therefore, we aggregate frames into a series of panoramic mosaics in the second view of the video player component, capturing a broader scope of what the performer sees at each timestep. 
We generate these panoramic mosaics by \revision{sampling frames from a \revision{temporal} window centered around the current timestep}. 
We then compute SIFT features for each frame~\cite{Lowe2004}, match \revision{them} using a Fast Library for Approximate Nearest Neighbors (FLANN)-based matcher~\cite{Muja2009FastAN}, and filter for valid matches by Lowe’s ratio test~\cite{Lowe2004} before warping and compositing the frames into a panoramic mosaic. 
We observe that these panoramic mosaics expand the view of the scene significantly, revealing objects within the field of view of the performer at a given timestep that were not captured by the main camera at that \revision{same} timestep (see Fig.~\ref{fig:stitching}).

%\erin{Since we're over the page limit, it may make sense to cut the entire paragraph below since it delves into the panoramic view in much more detail than is necessary for the overall design of ARGUS to make sense}
We note that in much of the existing literature on panoramic mosaics, the goal is to capture a seamless wider view of an (often static) scene at a single point in time. In these cases, previous works have endeavored to work around both in-scene and camera motion by excluding moving objects within the scene~\cite{HSU200481} or only addressing simple, slow camera panning motions~\cite{1544870}. 
When capturing video from an AR headset of a performer completing a task, however, unpredictable and rapid motion is not only unavoidable, but a valuable indicator of performer behavior. Therefore, our goal extends beyond the typical spatial expansion provided by a panoramic mosaic; we also aim to show how objects move around the complete scene over time, and how the object detection model performs over the given time range in order to facilitate both temporal and spatial analysis of a scene. 
%Moreover, w
We note that in our example task shown in Fig.~\ref{fig:stitching} (cooking), the \revision{performer} will often remain in the same position for many consecutive timesteps\revision{, consequently,} %and, therefore, 
the panoramic mosaic may not significantly expand the field of view at every timestep. 
Nevertheless, for tasks where the \revision{performer} traverses a larger area or turns their head in a wider range (and at timesteps where that behavior occurs in this task), the panoramic mosaic will significantly increase the portion of the scene shown at a given timestep.

%
%%%%%%%%%%%%%%%%%%%%%%%%%%%%%%%%%%%%%%%%%%%%%%%%%%%%%%%%%%%%%%%%%%%%%%%%%%
\begin{figure}[tbp]
    \centering
    \includegraphics[width=\linewidth]{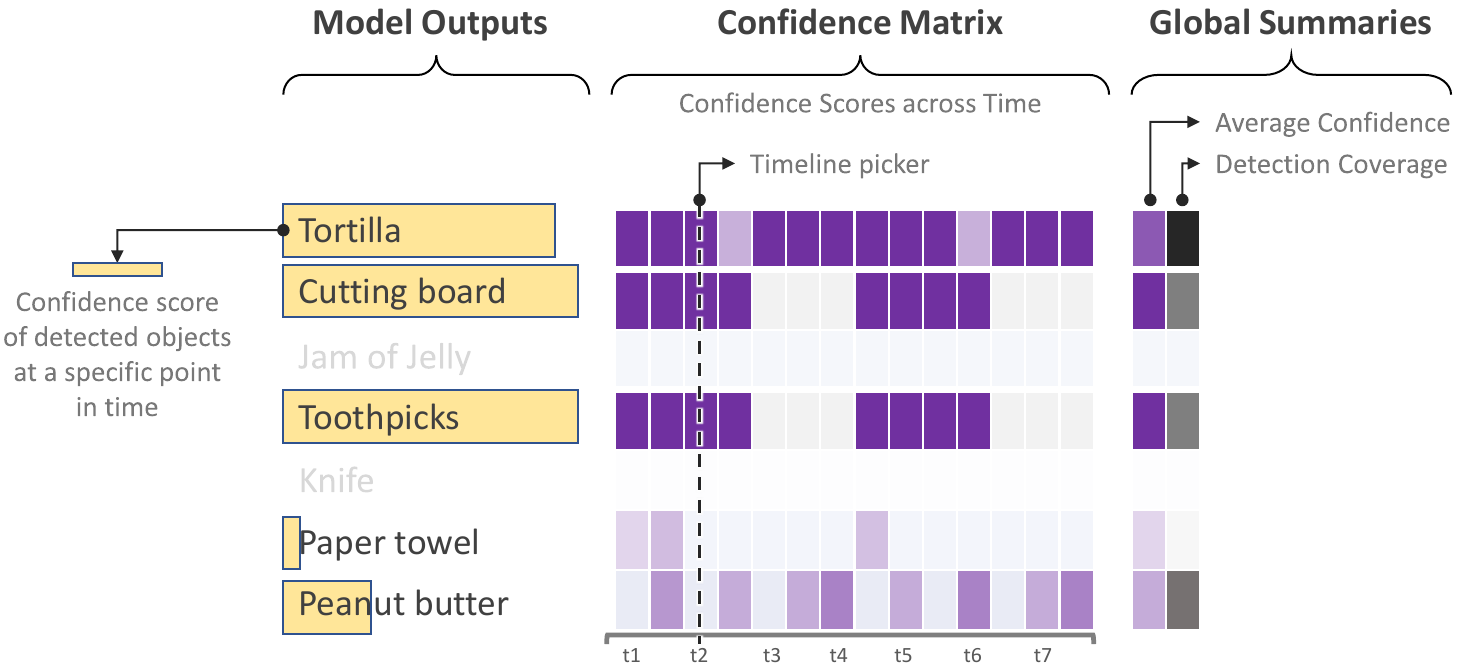}
    \vspace{-.2cm}
    \caption{\revision{Illustration of the Model Output Viewer applied to the analysis of a cooking recipe session. 
    %A confidence value of 0.1 is used as the detection threshold.
    To the left, the model outputs are listed vertically. 
    %The detected outputs' labels are highlighted with bars depicting the confidence level. 
    The bars depict the confidence \revision{score} of the detected outputs' labels at the specific time picked on the timeline. 
    In the middle, the temporal distribution of ML model output confidences across the whole session is displayed. 
    %Users can use the timeline picker to select a specific timeframe for further analysis.     
    To the right, two summaries are shown: the average confidence and detection coverage for each output across the entire session. 
    Color darkness is proportional to confidence value: \includegraphics[width=0.5cm,height=0.7em]{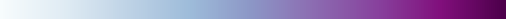}.
    }
    }
    \label{fig:temporal_viewer_visual_representation}
    \vspace{-.3cm}
\end{figure}
%%%%%%%%%%%%%%%%%%%%%%%%%%%%%%%%%%%%%%%%%%%%%%%%%%%%%%%%%%%%%%%%%%%%%%%%%%
%

\begin{comment}
\myparagraph{Temporal Controller} 
Includes a ``Play'' button that enables data playback, where the cursor is automatically advanced in real-time through the data. This component controls the video player and updates the \hyperref[fig:temporal_viewer_visual_representation]{Model Output Viewer} as well. Can be used to debug the models in a specific timeframe or to specify a data region for playback.
\end{comment}

\myparagraph{Model Output Viewer} During the debugging process of AR assistant models, the need for model output summaries is key to starting an analysis or evaluation. 
However, the temporal aspect inherent to these kinds of models makes this task more challenging since they often need to manage the sequence of actions or events chronologically. 
The \hyperref[fig:temporal_viewer_visual_representation]{Model Output Viewer} provides a summarization of the temporal distribution of the ML models outputs across the whole session (see Fig.~\ref{fig:systemoverview}\revision{(C3)}). 
This visualization is especially useful to find salient patterns, such as quick transitions between steps in step detection models, or to evaluate prediction consistency across time, allowing users to quickly have a global picture of the model behavior, something that could not be achieved by analyzing specific time frames.

As mentioned in Section \ref{subsec:modeling}, for AR assistive systems, the most relevant model outputs are the objects, actions, and steps. 
Once these model outputs are available, they are used to create the matrix visual representations for temporal model analysis. 
Fig.~\ref{fig:temporal_viewer_visual_representation} illustrates the \hyperref[fig:temporal_viewer_visual_representation]{Model Output Viewer}, where three main components are highlighted: the model outputs, the confidence matrix, and the global summaries. 
The \textit{Model output} view presents all the model outputs grouped by category. 
For example, as shown in Fig.~\ref{fig:systemoverview}\revision{(C3)}, there are three categories listed vertically: Objects, Actions, and Steps. 
The object, action, and step sections have multiple rows, each of them listing the model outputs for each category, e.g., the detected objects identified by the perception model. 
\textit{Confidence matrix}: The $x$-axis \revision{(or columns)} indicates the time, from 0 to the total duration of the session (video). 
%The matrix is filled with cells, where each cell represents the confidence score of the detected item at time \textit{t}. 
\revision{Each cell of t}he matrix \revision{is colored according to} the confidence score of the detected item at time \textit{t} (0\%\,\includegraphics[width=0.75cm,height=0.7em]{figs/imgs/bupu.png}\,100\%). 
%The matrix cell represents if the object or action has been detected at that point in time. 
If no action, object, or step is present, the matrix cell is left blank (white). %, otherwise, it is colored based on the confidence score. 
The total number of cells is proportional to the size of the session (seconds), and all cells are equal in height. 
Users can hover over the cells to see additional details.
% and have a width proportional to the number of cells divided by the chart width.
\textit{Global summaries}: The \hyperref[fig:temporal_viewer_visual_representation]{Model Output Viewer} also provides summaries of the average confidence and detection coverage for each row on the right side of the view so users can quickly evaluate them. 
The average confidence only takes the confidence value of detected objects, actions, or steps into account. 
Detection coverage refers to the total number of detections available for each model output (objects, actions, and steps). 

Even though the \hyperref[fig:systemoverview]{Temporal View} representation can provide a visual summary of the temporal distribution of the model output, details\revision{-}on\revision{-}demand functionalities remain crucial for debugging. 
The \hyperref[fig:temporal_viewer_visual_representation]{Model Output Viewer} allows users to do a focused analysis by letting them explore the model output results at specific points in time for further analysis. The user can use the temporal controller or the 3D viewer to do this selection. After this, all the objects, actions, and steps detected for that specific point in time that meets the confidence threshold are highlighted, as shown in Fig.~\ref{fig:systemoverview}\revision{(C3)}. 
Users can adjust the confidence threshold value using the slider to \revision{investigate} the object detection results. 
We also display object and action labels with bars depicting the confidence value for each label following guidelines of Felix et al.~\cite{Felix2018TakingWC} (see Fig.~\ref{fig:teaser}).\looseness=-1

\subsection{\revision{Implementation and Performance Details}}
\label{subsubsec:implementation}
The implementation of \systemname follows a set of constraints to allow for interactive query and rendering times. The backend supporting the rest API was written using Python and FastAPI~\cite{fast_api}. 
% All machine learning models were trained using PyTorch and Transformers \joao{Bea and Iran: any extra detail to add here?}. 
The ML models were trained and/or fine-tuned using PyTorch and serve predictions in real-time utilizing the same streaming protocol used by \systemname. 
The interface was structured as a \revision{dashboard-like} single-page application built with React~\cite{react} and TypeScript~\cite{typescript}. The visualization of 3D components uses Three.js~\cite{3js} and D3~\cite{d3}. % for the remaining. 
\revision{All the data consumed by \systemname \hyperref[fig:onlinedebugger]{online mode} comes from querying our Redis database, while the data available in the \hyperref[fig:systemoverview]{offline mode} comes from the data store in JSON format.} All the code is open source and hosted on GitHub\revision{~\cite{argus_2023}}. 

\revision{We have measured the latency of Microsoft's Windows Device Portal (part of their mixed reality capture \cite{msft_wdp}) at $\sim$1.3\,\revision{s} for streaming the main Hololens~2 camera, while ARGUS has a lower latency of $\sim$300\,ms. 
Currently, during online use, we save the various data streams at they get off the device. For the session in Section~\ref{subsec:spatial_case_study}, which takes 1:42\,min, the streamed point cloud has more than 10~M points, and it is highly redundant, since the same geometry is sampled over and over again. 
After the performer finishes a recording, we merge and downsample this data into a consolidated point cloud (see Sec.~\ref{subsubsec:spatial_view}), in this case with 70,000 points. We also create a voxel grid to generate the heat maps, which take 2.3\,s. After loading, all data is rendered in real-time.}

\section{Case Studies \& Discussion}
\label{sec:case_studies}
In this section, we present two case studies describing how model developers have made use of some of the available features. The section ends with feedback from domain experts who have used \systemname while developing AR task guidance software and a discussion of limitations.

\begin{figure}
    \centering
    \includegraphics[width=\linewidth]{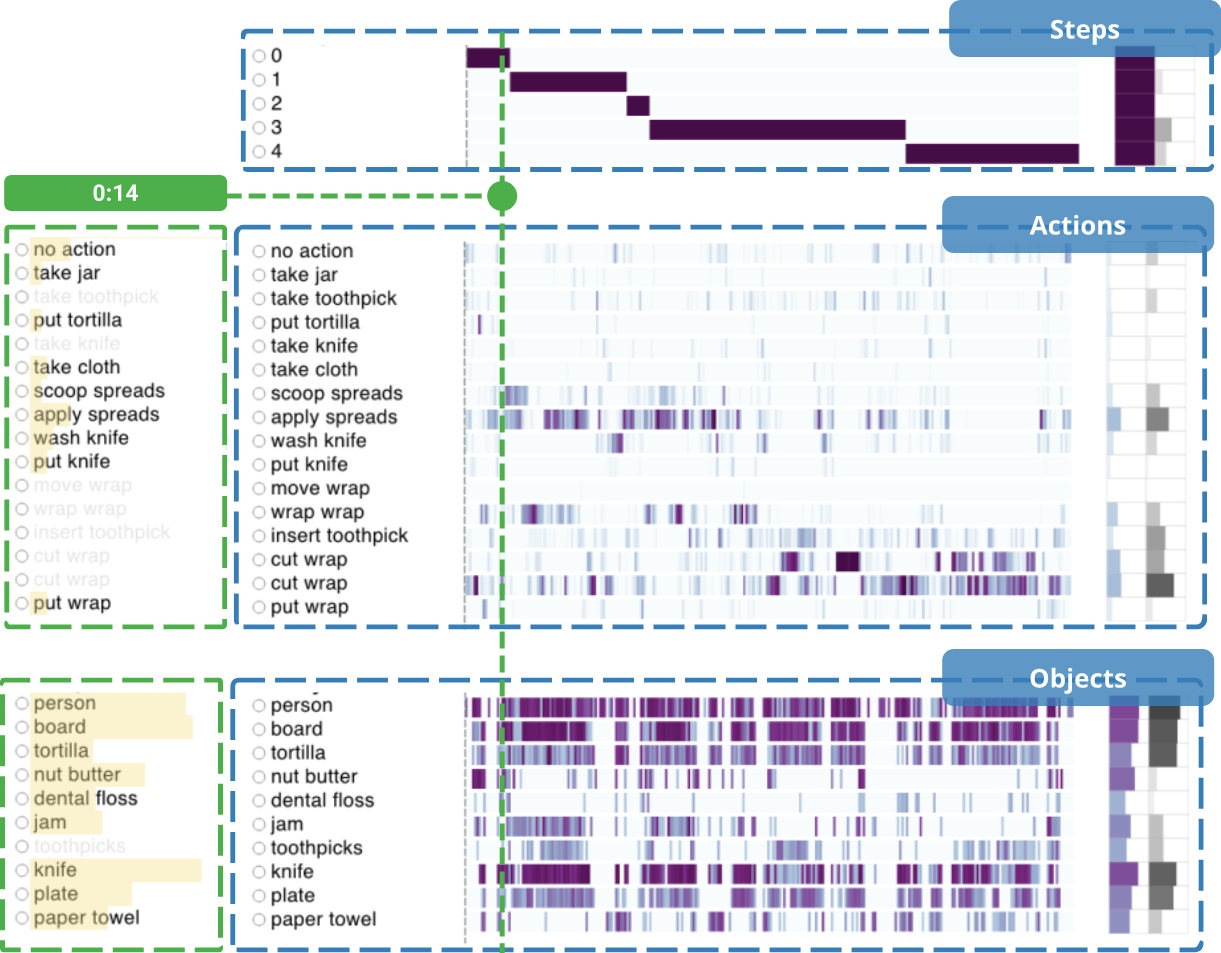}
    % \vspace{-.2cm}
    \captionof{figure}{\revision{Analysis of actions, objects, and steps in the Model Output Viewer. 
    Color darkness is proportional to confidence value (0\%\,\includegraphics[width=0.75cm,height=0.7em]{figs/imgs/bupu.png}\,100\%). 
    The confidence matrix and the average confidence views show that the confidence scores for objects are higher than actions. 
    The arrows show the confidence scores for actions and objects at minute 0:14 of the video. 
    The detection coverage view shows that some actions (e.g., \emph{take jar}) are rarely identified during the video.}}
    \label{fig:casestudy2_analysis}
    % \vspace{-.5cm}
\end{figure}
%

%
%
%%%%%%%%%%%%%%%%%%%%%%%%%%%%%%%%%%%%%%%%%%%%%%%%%%%%%%%%%%%%%%%%%%%%%%%%%
\begin{figure*}
    \centering
    \includegraphics[width=\linewidth]{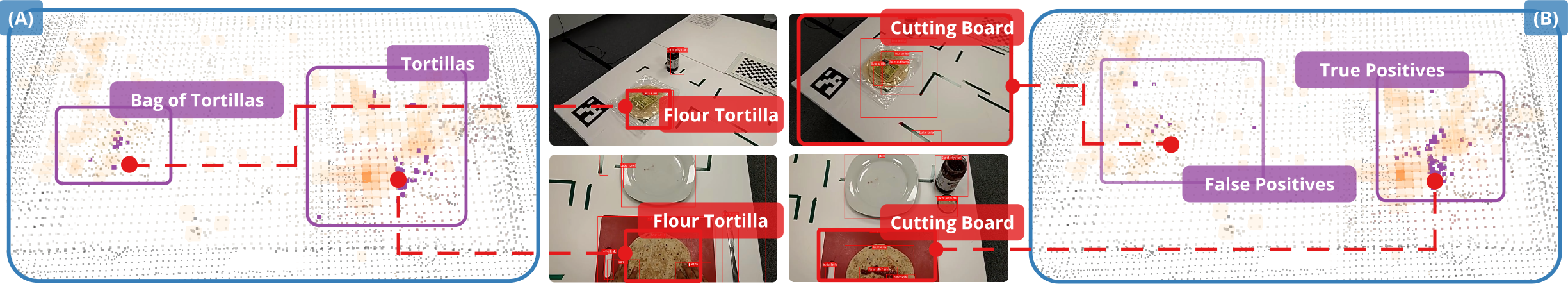}
    % \vspace{-.2cm} 
    \caption{\revision{Example of how the Spatial View can help users identify missing classes in the model's vocabulary or find clusters of false positives. (A) Points representing the 3D positions the object detection model identified as a "tortilla" during the session. Points on the left represent a bag of tortillas, while points on the right represent a single tortilla. (B) Points representing the 3D positions the object detection model identified as a "cutting board" during the session. The cluster on the left contains false positives, where the perception model generates wrong bounding boxes.
    }
    }
    \label{fig:modelheatmap}
    % \vspace{-.5cm}
\end{figure*}
%%%%%%%%%%%%%%%%%%%%%%%%%%%%%%%%%%%%%%%%%%%%%%%%%%%%%%%%%%%%%%%%%%%%%%%%%%
%
\subsection{Improving Step Transitions in Reasoning Module}\label{sec:improving_reasoning}

To showcase how the \hyperref[fig:temporal_viewer_visual_representation]{Model Output Viewer} supports the exploration and analysis of AI assistant model outputs (objects, actions, and steps)\revision{,} we describe how an ML engineer used this tool, the insights they gained, and how the reasoning module of the AI assistant, TIM, was improved through these insights. 
The ML engineer began by exploring the outputs of the reasoning and perception modules of a recorded session where a performer used TIM to follow a recipe\revision{~\cite{mitll_tasks}}.

\myparagraph{Analyzing step transitions} The visualization of the entire cooking session can help users find salient patterns, e.g., how the transitions between steps were carried out. The first repeated pattern identified by the ML engineer while using \systemname's \hyperref[fig:temporal_viewer_visual_representation]{Model Output Viewer} was the slow transition between steps (see Fig.~\ref{fig:casestudy2_analysis}). 
Investigating the ``Steps'' reveals that %. According to this figure, 
steps 1, 3, and 4 were performed over unexpectedly longer periods of time than steps 0 and 2. Also, the user noticed that the model only identified 5 out of 12 steps. Clearly, these two observations indicate that the reasoning module is making errors in identifying recipe steps. This visual summary of the \hyperref[fig:temporal_viewer_visual_representation]{Model Output Viewer} allows developers to quickly possess a global picture of model performance and assess errors. This could not be achieved \revision{as easily} without \systemname.

\myparagraph{Exploring detected objects} Under ``Objects'' in \revision{the} \hyperref[fig:temporal_viewer_visual_representation]{Model Output Viewer} (Fig.~\ref{fig:casestudy2_analysis}), the ML engineer noticed that the \revision{Detic} model identified most of the objects for the entirety of a recipe video. 
This \revision{is apparent from} the confidence matrix, where most rows are colored \revision{(}meaning an object was detected\revision{)}. 
The user also analyzed the confidence values of each detected object. 
For instance, at the 0:14 mark of the video, objects like \emph{board}, \emph{nut butter}, and \emph{knife} had high confidence values, indicated by the yellow background  (see zoomed-in views in Fig.~\ref{fig:casestudy2_analysis}). 
They could also see this trend in the ``Average Confidence'' column, which provides an average of confidence values throughout the video.

\myparagraph{Identifying missing actions} The ML engineer also analyzed the ``Actions'' section of the \hyperref[fig:temporal_viewer_visual_representation]{Model Output Viewer}. 
They noticed that some actions were rarely detected by the EgoVLP model. 
As we can see in the ``Detection Coverage'' column, actions like \emph{put knife}\revision{,} \emph{move wrap}\revision{,} and \emph{take cloth} were detected an unusually few number of times. 
This \revision{indicated} that it was difficult for EgoVLP to detect those actions. They also noticed that the confidence values for the actions were much lower than the ones for objects. 
As \revision{is} visible in Fig.~\ref{fig:casestudy2_analysis}, the predominant color during the whole session was light purple, which represents low confidence in detecting the actions. 
Also, at time 0:14 of the video, the confidence values for ``scoop spreads'' and ``apply spreads'' were low. 
In the ``Average Confidence'' column, we can see that actions such as ``wash knife'' and ``insert toothpick'' had approximately 30\% confidence. 
This information led the ML engineer to hypothesize that a decrease in the confidence threshold might be necessary to recognize steps effectively.
\revision{The visualizations provided by \systemname also help to investigate whether lowering the threshold would lead to false positives in the step recognition.}

\myparagraph{Using insights to improve the reasoning module} After the analysis, the ML engineer modified the reasoning module to handle actions with low confidence. The reasoning module defaults to selecting actions with greater than 70\% confidence. 
The ML engineer used the confidence slider of \hyperref[fig:temporal_viewer_visual_representation]{Model Output Viewer} to tune this value. 
The most promising value they found was 30\%. 
They reran the new version of the reasoning module for the same video. 
As shown in Table~\ref{tab:steps_accuracy}, the step estimation accuracy increased for every step and from 35\% to 73\% overall. 
Also, this new version was able to identify 8 out of 12 steps, while the previous version only identified 4 out of 12 steps. 

\myparagraph{Interpreting the new results} The \hyperref[fig:temporal_viewer_visual_representation]{Model Output Viewer} was also useful for the ML engineer to understand why the reasoning module failed to recognize some steps. 
As we can see in Table~\ref{tab:steps_accuracy}, steps 4, 5, and 8 were not recognized. For instance, step 5 (``Roll the tortilla from one end to the other into a log shape'') is directly related to the action ``move wrap'', and this action was not identified at all during the entire session (see ``Detection Coverage'' column). Since this action is necessary for step 5, it was not identified by the reasoning module.

\begin{table}[t]
\caption{Accuracy of the old and new version of the reasoning module for recognizing the steps of the recipe.}
\centering
\scriptsize
\setlength{\tabcolsep}{3.1pt} % Default value: 6pt
\begin{tabular}{lrrrrrrrrrrrrr}%{llllllllllllll}
\toprule
Version  & S0 & S1 & S2 & S3 & S4 & S5 & S6 & S7 & S8 & S9 & S10 & S11& Total \\
\midrule
Old  & 1.0 & 0.9 & 0.3 & 0.9 & 0 & 0   & 0   & 0   & 0   & 0  & 0 & 0   & 0.35   \\
New & 1.0 & 1.0 & 0.5 & 1.0 & 0 & 0   & 0.4   & 0.9   & 0   & 1.0   & 1.0  & 0 & 0.73   \\
\bottomrule  
\end{tabular}
\label{tab:steps_accuracy}
\end{table}

\subsection{\revision{Using Spatial Features to Explain Failures}}\label{subsec:spatial_case_study}

Although the \hyperref[fig:systemoverview]{Temporal View} can help users uncover undesired patterns in model performance, it does not paint the full picture of the situation, \revision{as} model failures might be related to spatial characteristics or \revision{performer} behavior. 
In this case study, we show %case 
how the \hyperref[fig:systemoverview]{Spatial View} can provide deeper insights into both reasoning and perception models by assisting users in finding regions where the perception models underperform and to correlate \revision{performer} behavior with reasoning outputs.

A very common way to assess the quality of perception models is by checking the spatial distribution of static objects. In other words, the physical objects captured by the headset cameras can generally be classified as either static objects (objects that will likely not move) or dynamic objects. This classification can help users quickly identify regions where the perception model fails by detecting objects not expected to move throughout a recording session. This case study highlights how this sanity check becomes trivial in \systemname.

We start the exploration by first using the \hyperref[fig:systemoverview]{Data Manager} to load the parts of a recording where the perception model underperformed. 
Once a recording from this period is loaded, we can %immediately 
use the \hyperref[fig:systemoverview]{Spatial View} to find regions of the space \revision{where} the performer was interacting. 
Fig.~\ref{fig:systemoverview} shows points of \revision{performer} positions (blue) and gaze projection (orange). 
%After a quick 
\revision{During an initial} inspection, the user can quickly recognize three darker regions projected in the world's point cloud. 
%First, the rightmost region represents the time during which the performer was interacting with \systemname in \hyperref[fig:onlinedebugger]{online mode} to start the recording. 
%Secondly, two regions on the desk appear as darker regions in the heatmap. 
\revision{The rightmost region represents the time during which the performer was interacting with \systemname in \hyperref[fig:onlinedebugger]{online mode} to start the recording, while the other two regions are on the desk.}
The user then hovers the mouse over the points on the 3D point cloud representing the gaze projection on the world point cloud to look at the corresponding video frames in \hyperref[fig:systemoverview]{Temporal View}. 
This interaction reveals \revision{that} the left region contains the ingredients for the recipe, while the actions (e.g., \revision{``}spreading jelly on the tortilla\revision{''}) happen on the right side. 
By highlighting the heatmaps only, it becomes clear that the performer spent most of their time looking at the right side of the table (darker region), meaning the performer spent more time executing actions than selecting ingredients.    
With this understanding of the spatial distribution of the performer's attention, the user can infer that the model outputs high confidence values for \emph{tortilla} and \emph{cutting board} throughout the entire session as shown in Fig.~\ref{fig:casestudy2_analysis}, which makes the user question the validity of the output.
Then, the user displays the 3D point cloud denoting the 3D positions of tortillas shown in \revision{Fig. \ref{fig:modelheatmap}(A)}. Two clusters show up, allowing the user to look at the corresponding frames, realizing that the left cluster represents a bag of tortillas while the other is the tortilla used for the recipe. This process highlights the need for a more comprehensive class vocabulary able to represent both tortillas and bags of tortillas.
Following that, the same process is conducted for the \emph{cutting board} class, and a similar pattern with two clusters \revision{arises} \revision{(see Fig.~\ref{fig:modelheatmap}(B))}. 
%With the cutting board being one of the static objects in the recording, the user can quickly realize that one of the clusters may be representing a model failure.
\revision{Since} the cutting board \revision{was a} static object in the recording, the user can quickly realize that one of the clusters may be representing a model failure. 
The corresponding video frames selected interactively \revision{confirm} that the left cluster contains only false negatives.
Lastly, inspecting the bounding boxes rendered on the video frames (see Fig. \ref{fig:modelheatmap}) gives the user more detailed information about the error. In this case, the user sees that the model is generating bounding boxes covering almost the entire field of view of the \revision{performer}.

\vspace{-0.05in}

\subsection{Expert Feedback}
The \hyperref[fig:temporal_viewer_visual_representation]{Model Output Viewer} provides a visualization of object and action detections with model confidence levels. Even if a model performs very well on an offline evaluation dataset, when deployed in real-time\revision{,} it will inevitably be presented with previously unseen conditions such as room lighting, skin pigmentation, or object angle. This is known as the ``domain-shift'' problem \cite{zhou2021domain}, where a model fails to perform when presented with data %that is 
not well represented in its offline evaluation dataset. 
\systemname streamlines real-time deployment, and its \hyperref[fig:temporal_viewer_visual_representation]{Model Output Viewer} \revision{enables} %allows for 
the evaluation of model confidence in a virtually unconstrained domain. This sheds light on which conditions the models perform best, and informs how model robustness could be enhanced by expanding data with new collection or augmentation strategies. 

% \systemname also facilitates the collection of high-quality training data, allowing for the data collection performers to assess the data quality as they are performing the task. It provides convenient tools to control recordings as well as detect errors early, before 

% ARGUS allprovides machine learning researchers with several tools

\systemname is also useful for scenarios where multiple information sources must be analyzed at the same time. For TIM's reasoning module, which consumes multiple inputs in parallel (e.g., the detected objects/actions and its confidence scores), \systemname's visualizations allow the user to understand the reasons the system made the predictions and under what circumstances it succeeded/failed. As shown in Sec.~\ref{sec:improving_reasoning}, this tool helped the ML engineer to improve the system.

As ML models develop new capabilities and produce richer representations, it becomes increasingly important to develop scalable visualizations of those outputs. 
Conventionally, ML engineers either log outputs to the terminal or use drawing libraries to bake the predictions on top of the video. 
However, there is limited real estate when drawing on a video, and often the predictions and their associated text make it difficult to view the underlying image frames. 
%Instead, web technology provides a level of interactivity that is difficult and fickle to get by other means and it allows you to selectively show relevant information while allowing the user to change the view and granularity of that information to suit their needs. 
\revision{In contrast, \systemname provides a high level of interactivity, which} allows it to selectively \revision{visualize} relevant information while allowing the user to change the view and granularity of \revision{these} information to suit their needs.
% \systemname does this well with it's
Additionally, being able to contextualize and explore ML model outputs in 3D can lead to a better understanding of how model outputs can change based on the perspective, and spatially grounds the predictions for an entire recording in a single view. 
Overall, tools like \systemname drastically lighten the visualization load placed on ML engineers and provide a convenient tool for understanding their models.

\myparagraph{Limitations} While useful for exploration of spatiotemporal data captured by an intelligent assistant, \systemname \revision{needs more robust data processing algorithms.}
\revision{For instance,} in sessions where the \revision{performer}'s hands are recurrently in the field of view of the headset camera, the point cloud generation process captures and transforms it into points of the world space, resulting in potential noise that does not represent the physical environment. To overcome this problem, we review recordings with noisy point clouds and define bounding boxes representing regions where these noisy points must be excluded from the final rendering. We plan to explore methods~\cite{miknis2015near,duan2021low} to automatically remove point cloud noise during run-time acquisition.

%\vspace{-.1in}

\section{Conclusion \& Future Work}
\label{sec:conclusion}
We presented \systemname, an interactive visual analytics system that empowers developers of intelligent assistive AR systems \revision{to} seamless\revision{ly analyze} complex datasets, created by integrating multiple data streams at different scales, dimensions, and formats acquired during performance time. Furthermore, through interactive and well-integrated \hyperref[fig:systemoverview]{spatial} and \hyperref[fig:systemoverview]{temporal} visualization widgets, \revision{it} allows for retrospective analysis and debugging of historical data generated by ML models for AI assistants. 

We envision \systemname to unlock several avenues for future research connecting human-computer interaction, visualization, and machine learning communities revolving around the goal of developing better and more reliable AR intelligent systems. 
\revision{In the future, we intend to conduct a deeper evaluation of our system's performance metrics (e.g. rendering times, stream latency).} 
We also plan to explore how to extend the system to support the comparison of sessions of multiple \revision{performers}. 
This includes the data and model outputs and will require registration of the point clouds. 
User-generated data acquisition (annotation) and integrated AI techniques during exploration time (segmentation and model training based on the annotated data) are other fronts we would like to cover. 
Since our \hyperref[fig:systemoverview]{Temporal} and \hyperref[fig:systemoverview]{Spatial} Views allow users to explore data and output models across the entire session, adding annotation capabilities is a natural next step.
\revision{Furthermore, we want to investigate privacy-preserving methods for storing and streaming the collected data, similar to ones that have been proposed, e.g., for eye-tracking data~\cite{bozkir_privacy_2021,davidjohn_privacy_2021}, to prevent performer identification.}
%% if specified like this the section will be ommitted in review mode
\acknowledgments{%
We would like to thank Jay Jatin Sutaria and Fabio Felix Dias for help with the development of \systemname. This work was supported by the DARPA PTG program.  Any opinions, findings, and conclusions or recommendations expressed in this material are those of the authors and do not necessarily reflect the views of DARPA.
Erin McGowan is funded by an NYU Tandon Future Leader Fellowship.}
% \section*{Acknowledgements}

\bibliographystyle{abbrv-doi-hyperref}

\bibliography{references.bib}

\begin{thebibliography}{10}

\bibitem{aigner_visualization_2011}
W.~Aigner, S.~Miksch, H.~Schumann, and C.~Tominski.
\newblock {\em Visualization of {Time}-{Oriented} {Data}}.
\newblock Human–{Computer} {Interaction} {Series}. Springer-Verlag, London,
  2011. \href{https://doi.org/10.1007/978-0-85729-079-3}
{doi: {{%
10\hspace{.1pt}\discretionary{.}{%
}{.}\hspace{.4pt}1007\discretionary{/}{%
}{/}978\discretionary{%
}{-}{-}0\discretionary{%
}{-}{-}85729\discretionary{%
}{-}{-}079\discretionary{%
}{-}{-}3}}}


\bibitem{argus_2023}
{ARGUS}.
\newblock Augmented reality guidance and user-modeling system.
\newblock \url{https://github.com/VIDA-NYU/ARGUS}, 2023.

\bibitem{baum1966statistical}
L.~E. Baum and T.~Petrie.
\newblock Statistical inference for probabilistic functions of finite state
  markov chains.
\newblock {\em The annals of mathematical statistics}, 37(6):1554--1563, 1966.

\bibitem{beams_evaluation_2022}
R.~Beams, E.~Brown, W.-C. Cheng, J.~S. Joyner, A.~S. Kim, K.~Kontson,
  D.~Amiras, T.~Baeuerle, W.~Greenleaf, R.~J. Grossmann, A.~Gupta, C.~Hamilton,
  H.~Hua, T.~T. Huynh, C.~Leuze, S.~B. Murthi, J.~Penczek, J.~Silva,
  B.~Spiegel, A.~Varshney, and A.~Badano.
\newblock Evaluation {Challenges} for the {Application} of {Extended} {Reality}
  {Devices} in {Medicine}.
\newblock {\em Journal of Digital Imaging}, 35(5):1409--1418, 2022.
  \href{https://doi.org/10.1007/s10278-022-00622-x}
{doi: {{%
10\hspace{.1pt}\discretionary{.}{%
}{.}\hspace{.4pt}1007\discretionary{/}{%
}{/}s10278\discretionary{%
}{-}{-}022\discretionary{%
}{-}{-}00622\discretionary{%
}{-}{-}x}}}


\bibitem{becher_situated_2022}
M.~Becher, D.~Herr, C.~Müller, K.~Kurzhals, G.~Reina, L.~Wagner, T.~Ertl, and
  D.~Weiskopf.
\newblock Situated {Visual} {Analysis} and {Live} {Monitoring} for
  {Manufacturing}.
\newblock {\em IEEE Computer Graphics and Applications}, 42(2):33--44, 2022.
  \href{https://doi.org/10.1109/MCG.2022.3157961}
{doi: {{%
10\hspace{.1pt}\discretionary{.}{%
}{.}\hspace{.4pt}1109\discretionary{/}{%
}{/}MCG\hspace{.1pt}\discretionary{.}{%
}{.}\hspace{.4pt}2022\hspace{.1pt}\discretionary{.}{%
}{.}\hspace{.4pt}3157961}}}


\bibitem{psi}
D.~Bohus, S.~Andrist, A.~Feniello, N.~Saw, M.~Jalobeanu, P.~Sweeney, A.~L.
  Thompson, and E.~Horvitz.
\newblock Platform for situated intelligence.
\newblock {\em CoRR}, abs/2103.15975, 2021.

\bibitem{d3}
M.~Bostock.
\newblock D3.js.
\newblock \url{https://d3js.org/}.

\bibitem{bozkir_privacy_2021}
E.~Bozkir, O.~Günlü, W.~Fuhl, R.~F. Schaefer, and E.~Kasneci.
\newblock Differential privacy for eye tracking with temporal correlations.
\newblock {\em PLOS ONE}, 16(8):1--22, 2021.
  \href{https://doi.org/10.1371/journal.pone.0255979}
{doi: {{%
10\hspace{.1pt}\discretionary{.}{%
}{.}\hspace{.4pt}1371\discretionary{/}{%
}{/}journal\hspace{.1pt}\discretionary{.}{%
}{.}\hspace{.4pt}pone\hspace{.1pt}\discretionary{.}{%
}{.}\hspace{.4pt}0255979}}}


\bibitem{caudell_augmented_1992}
T.~Caudell and D.~Mizell.
\newblock Augmented reality: an application of heads-up display technology to
  manual manufacturing processes.
\newblock In {\em Proceedings of the {Twenty}-{Fifth} {Hawaii} {International}
  {Conference} on {System} {Sciences}}, vol.~ii, pp. 659--669, 1992.
  \href{https://doi.org/10.1109/HICSS.1992.183317}
{doi: {{%
10\hspace{.1pt}\discretionary{.}{%
}{.}\hspace{.4pt}1109\discretionary{/}{%
}{/}HICSS\hspace{.1pt}\discretionary{.}{%
}{.}\hspace{.4pt}1992\hspace{.1pt}\discretionary{.}{%
}{.}\hspace{.4pt}183317}}}


\bibitem{ptg_site}
{DARPA}.
\newblock Perceptually-enabled task guidance {(PTG)}.
\newblock
  \url{https://www.darpa.mil/program/perceptually-enabled-task-guidance}.

\bibitem{davidjohn_privacy_2021}
B.~David-John, D.~Hosfelt, K.~Butler, and E.~Jain.
\newblock A privacy-preserving approach to streaming eye-tracking data.
\newblock {\em IEEE Transactions on Visualization and Computer Graphics},
  27(5):2555--2565, 2021. \href{https://doi.org/10.1109/TVCG.2021.3067787}
{doi: {{%
10\hspace{.1pt}\discretionary{.}{%
}{.}\hspace{.4pt}1109\discretionary{/}{%
}{/}TVCG\hspace{.1pt}\discretionary{.}{%
}{.}\hspace{.4pt}2021\hspace{.1pt}\discretionary{.}{%
}{.}\hspace{.4pt}3067787}}}


\bibitem{duan2021low}
Y.~Duan, C.~Yang, H.~Chen, W.~Yan, and H.~Li.
\newblock Low-complexity point cloud denoising for lidar by pca-based dimension
  reduction.
\newblock {\em Optics Communications}, 482:126567, 2021.

\bibitem{feichtenhofer2019slowfast}
C.~Feichtenhofer, H.~Fan, J.~Malik, and K.~He.
\newblock Slowfast networks for video recognition.
\newblock In {\em Proceedings of the IEEE/CVF international conference on
  computer vision}, pp. 6202--6211, 2019.

\bibitem{Felix2018TakingWC}
C.~Felix, S.~L. Franconeri, and E.~Bertini.
\newblock Taking word clouds apart: An empirical investigation of the design
  space for keyword summaries.
\newblock {\em IEEE Transactions on Visualization and Computer Graphics},
  24:657--666, 2018.

\bibitem{fernandez_del_amo_augmented_2018}
I.~Fernández~del Amo, J.~A. Erkoyuncu, R.~Roy, and S.~Wilding.
\newblock Augmented {Reality} in {Maintenance}: {An} information-centred design
  framework.
\newblock {\em Procedia Manufacturing}, 19:148--155, 2018.
  \href{https://doi.org/10.1016/j.promfg.2018.01.021}
{doi: {{%
10\hspace{.1pt}\discretionary{.}{%
}{.}\hspace{.4pt}1016\discretionary{/}{%
}{/}j\hspace{.1pt}\discretionary{.}{%
}{.}\hspace{.4pt}promfg\hspace{.1pt}\discretionary{.}{%
}{.}\hspace{.4pt}2018\hspace{.1pt}\discretionary{.}{%
}{.}\hspace{.4pt}01\hspace{.1pt}\discretionary{.}{%
}{.}\hspace{.4pt}021}}}


\bibitem{fleck_ragrug_2022}
P.~Fleck, A.~Sousa~Calepso, S.~Hubenschmid, M.~Sedlmair, and D.~Schmalstieg.
\newblock {RagRug}: {A} {Toolkit} for {Situated} {Analytics}.
\newblock {\em IEEE Transactions on Visualization and Computer Graphics}, pp.
  1--1, 2022. \href{https://doi.org/10.1109/TVCG.2022.3157058}
{doi: {{%
10\hspace{.1pt}\discretionary{.}{%
}{.}\hspace{.4pt}1109\discretionary{/}{%
}{/}TVCG\hspace{.1pt}\discretionary{.}{%
}{.}\hspace{.4pt}2022\hspace{.1pt}\discretionary{.}{%
}{.}\hspace{.4pt}3157058}}}


\bibitem{foxglove_2023}
Foxglove.
\newblock Foxglove - {Visualizing} and debugging your robotics data.

\bibitem{funk2015using}
M.~Funk, S.~Mayer, and A.~Schmidt.
\newblock Using in-situ projection to support cognitively impaired workers at
  the workplace.
\newblock In {\em Proceedings of the 17th international ACM SIGACCESS
  conference on Computers \& accessibility}, pp. 185--192, 2015.

\bibitem{girdhar2022omnivore}
R.~Girdhar, M.~Singh, N.~Ravi, L.~van~der Maaten, A.~Joulin, and I.~Misra.
\newblock Omnivore: A single model for many visual modalities.
\newblock In {\em Proceedings of the IEEE/CVF Conference on Computer Vision and
  Pattern Recognition}, pp. 16102--16112, 2022.

\bibitem{henderson_exploring_2011}
S.~Henderson and S.~Feiner.
\newblock Exploring the {Benefits} of {Augmented} {Reality} {Documentation} for
  {Maintenance} and {Repair}.
\newblock {\em IEEE Transactions on Visualization and Computer Graphics},
  17(10):1355--1368, 2011. \href{https://doi.org/10.1109/TVCG.2010.245}
{doi: {{%
10\hspace{.1pt}\discretionary{.}{%
}{.}\hspace{.4pt}1109\discretionary{/}{%
}{/}TVCG\hspace{.1pt}\discretionary{.}{%
}{.}\hspace{.4pt}2010\hspace{.1pt}\discretionary{.}{%
}{.}\hspace{.4pt}245}}}


\bibitem{HSU200481}
C.-T. Hsu and Y.-C. Tsan.
\newblock Mosaics of video sequences with moving objects.
\newblock {\em Signal Processing: Image Communication}, 19(1):81--98, 2004.
  \href{https://doi.org/10.1016/j.image.2003.10.001}
{doi: {{%
10\hspace{.1pt}\discretionary{.}{%
}{.}\hspace{.4pt}1016\discretionary{/}{%
}{/}j\hspace{.1pt}\discretionary{.}{%
}{.}\hspace{.4pt}image\hspace{.1pt}\discretionary{.}{%
}{.}\hspace{.4pt}2003\hspace{.1pt}\discretionary{.}{%
}{.}\hspace{.4pt}10\hspace{.1pt}\discretionary{.}{%
}{.}\hspace{.4pt}001}}}


\bibitem{jiang_hololens-based_2020}
T.~Jiang, D.~Yu, Y.~Wang, T.~Zan, S.~Wang, and Q.~Li.
\newblock {HoloLens}-{Based} {Vascular} {Localization} {System}.
\newblock {\em Journal of Medical Internet Research}, 22(4):e16852, Apr. 2020.
  \href{https://doi.org/10.2196/16852}
{doi: {{%
10\hspace{.1pt}\discretionary{.}{%
}{.}\hspace{.4pt}2196\discretionary{/}{%
}{/}16852}}}


\bibitem{kazakos2021slow}
E.~Kazakos, A.~Nagrani, A.~Zisserman, and D.~Damen.
\newblock Slow-fast auditory streams for audio recognition.
\newblock In {\em ICASSP 2021-2021 IEEE International Conference on Acoustics,
  Speech and Signal Processing (ICASSP)}, pp. 855--859. IEEE, 2021.

\bibitem{kehrer_visualization_2013}
J.~Kehrer and H.~Hauser.
\newblock Visualization and {Visual} {Analysis} of {Multifaceted} {Scientific}
  {Data}: {A} {Survey}.
\newblock {\em IEEE Transactions on Visualization and Computer Graphics},
  19(3):495--513, 2013. \href{https://doi.org/10.1109/TVCG.2012.110}
{doi: {{%
10\hspace{.1pt}\discretionary{.}{%
}{.}\hspace{.4pt}1109\discretionary{/}{%
}{/}TVCG\hspace{.1pt}\discretionary{.}{%
}{.}\hspace{.4pt}2012\hspace{.1pt}\discretionary{.}{%
}{.}\hspace{.4pt}110}}}


\bibitem{kim_does_2018}
K.~Kim, L.~Boelling, S.~Haesler, J.~Bailenson, G.~Bruder, and G.~F. Welch.
\newblock Does a {Digital} {Assistant} {Need} a {Body}? {The} {Influence} of
  {Visual} {Embodiment} and {Social} {Behavior} on the {Perception} of
  {Intelligent} {Virtual} {Agents} in {AR}.
\newblock In {\em 2018 {IEEE} {International} {Symposium} on {Mixed} and
  {Augmented} {Reality} ({ISMAR})}, pp. 105--114, 2018.
\newblock ISSN: 1554-7868. \href{https://doi.org/10.1109/ISMAR.2018.00039}
{doi: {{%
10\hspace{.1pt}\discretionary{.}{%
}{.}\hspace{.4pt}1109\discretionary{/}{%
}{/}ISMAR\hspace{.1pt}\discretionary{.}{%
}{.}\hspace{.4pt}2018\hspace{.1pt}\discretionary{.}{%
}{.}\hspace{.4pt}00039}}}


\bibitem{lin2020recipe}
A.~Lin, S.~Rao, A.~Celikyilmaz, E.~Nouri, C.~Brockett, D.~Dey, and W.~B. Dolan.
\newblock A recipe for creating multimodal aligned datasets for sequential
  tasks.
\newblock In {\em Proceedings of the 58th Annual Meeting of the Association for
  Computational Linguistics}, pp. 4871--4884, 2020.

\bibitem{liu_smart_2018}
C.-F. Liu and P.-Y. Chiang.
\newblock Smart glasses based intelligent trainer for factory new recruits.
\newblock In {\em Proceedings of the 20th {International} {Conference} on
  {Human}-{Computer} {Interaction} with {Mobile} {Devices} and {Services}
  {Adjunct}}, {MobileHCI} '18, pp. 395--399. Association for Computing
  Machinery, 2018. \href{https://doi.org/10.1145/3236112.3236174}
{doi: {{%
10\hspace{.1pt}\discretionary{.}{%
}{.}\hspace{.4pt}1145\discretionary{/}{%
}{/}3236112\hspace{.1pt}\discretionary{.}{%
}{.}\hspace{.4pt}3236174}}}


\bibitem{liu_visualizing_2017}
S.~Liu, D.~Maljovec, B.~Wang, P.-T. Bremer, and V.~Pascucci.
\newblock Visualizing {High}-{Dimensional} {Data}: {Advances} in the {Past}
  {Decade}.
\newblock {\em IEEE Transactions on Visualization and Computer Graphics},
  23(3):1249--1268, 2017. \href{https://doi.org/10.1109/TVCG.2016.2640960}
{doi: {{%
10\hspace{.1pt}\discretionary{.}{%
}{.}\hspace{.4pt}1109\discretionary{/}{%
}{/}TVCG\hspace{.1pt}\discretionary{.}{%
}{.}\hspace{.4pt}2016\hspace{.1pt}\discretionary{.}{%
}{.}\hspace{.4pt}2640960}}}


\bibitem{Lowe2004}
D.~G. Lowe.
\newblock Distinctive image features from scale-invariant keypoints.
\newblock {\em International Journal of Computer Vision}, 60(2):91--110, Nov.
  2004. \href{https://doi.org/10.1023/b:visi.0000029664.99615.94}
{doi: {{%
10\hspace{.1pt}\discretionary{.}{%
}{.}\hspace{.4pt}1023\discretionary{/}{%
}{/}b\discretionary{:}{%
}{:}visi\hspace{.1pt}\discretionary{.}{%
}{.}\hspace{.4pt}0000029664\hspace{.1pt}\discretionary{.}{%
}{.}\hspace{.4pt}99615\hspace{.1pt}\discretionary{.}{%
}{.}\hspace{.4pt}94}}}


\bibitem{marriott_immersive_2018}
K.~Marriott, F.~Schreiber, T.~Dwyer, K.~Klein, N.~H. Riche, T.~Itoh,
  W.~Stuerzlinger, and B.~H. Thomas, eds.
\newblock {\em Immersive {Analytics}}.
\newblock Springer International Publishing, 2018.
  \href{https://doi.org/10.1007/978-3-030-01388-2}
{doi: {{%
10\hspace{.1pt}\discretionary{.}{%
}{.}\hspace{.4pt}1007\discretionary{/}{%
}{/}978\discretionary{%
}{-}{-}3\discretionary{%
}{-}{-}030\discretionary{%
}{-}{-}01388\discretionary{%
}{-}{-}2}}}


\bibitem{typescript}
Microsoft.
\newblock Typescript.
\newblock \url{https://www.typescriptlang.org/}.

\bibitem{msft_wdp}
Microsoft.
\newblock Using the windows device portal.
\newblock
  \url{https://learn.microsoft.com/en-us/windows/mixed-reality/develop/advanced-concepts/using-the-windows-device-portal},
  2022.

\bibitem{miknis2015near}
M.~Miknis, R.~Davies, P.~Plassmann, and A.~Ware.
\newblock Near real-time point cloud processing using the pcl.
\newblock In {\em 2015 International Conference on Systems, Signals and Image
  Processing (IWSSIP)}, pp. 153--156. IEEE, 2015.

\bibitem{milgram_taxonomy_1994}
P.~Milgram and F.~Kishino.
\newblock A {Taxonomy} of {Mixed} {Reality} {Visual} {Displays}.
\newblock {\em IEICE Transactions on Information Systems}, E77-D, 1994.

\bibitem{mitll_tasks}
{MIT Lincoln Laboratory}.
\newblock {PTG} evaluation tasks vol. 1.
\newblock \url{TBD}, 2022.

\bibitem{Muja2009FastAN}
M.~Muja and D.~G. Lowe.
\newblock Fast approximate nearest neighbors with automatic algorithm
  configuration.
\newblock In {\em International Conference on Computer Vision Theory and
  Applications}, 2009.

\bibitem{nair2023truth}
V.~Nair, L.~Rosenberg, J.~F. O'Brien, and D.~Song.
\newblock Truth in motion: The unprecedented risks and opportunities of
  extended reality motion data.
\newblock \url{https://arxiv.org/abs/2306.06459}, 2023.

\bibitem{nijholt_towards_2022}
A.~Nijholt.
\newblock Towards {Social} {Companions} in {Augmented} {Reality}: {Vision} and
  {Challenges}.
\newblock In N.~A. Streitz and S.~Konomi, eds., {\em Distributed, {Ambient} and
  {Pervasive} {Interactions}. {Smart} {Living}, {Learning}, {Well}-being and
  {Health}, {Art} and {Creativity}}, Lecture {Notes} in {Computer} {Science},
  pp. 304--319. Springer International Publishing, 2022.
  \href{https://doi.org/10.1007/978-3-031-05431-0_21}
{doi: {{%
10\hspace{.1pt}\discretionary{.}{%
}{.}\hspace{.4pt}1007\discretionary{/}{%
}{/}978\discretionary{%
}{-}{-}3\discretionary{%
}{-}{-}031\discretionary{%
}{-}{-}05431\discretionary{%
}{-}{-}0\_21}}}


\bibitem{pase2012ethical}
S.~Pase.
\newblock Ethical considerations in augmented reality applications.
\newblock In {\em Proceedings of the International Conference on e-Learning,
  e-Business, Enterprise Information Systems, and e-Government (EEE)}, p.~1.
  The Steering Committee of The World Congress in Computer Science, Computer
  Engineering and Applied Computing (WorldComp), 2012.

\bibitem{puladi_augmented_2022}
B.~Puladi, M.~Ooms, M.~Bellgardt, M.~Cesov, M.~Lipprandt, S.~Raith, F.~Peters,
  S.~C. Möhlhenrich, A.~Prescher, F.~Hölzle, T.~W. Kuhlen, and A.~Modabber.
\newblock Augmented {Reality}-{Based} {Surgery} on the {Human} {Cadaver}
  {Using} a {New} {Generation} of {Optical} {Head}-{Mounted} {Displays}.
\newblock {\em JMIR Serious Games}, 10(2):e34781, 2022.
  \href{https://doi.org/10.2196/34781}
{doi: {{%
10\hspace{.1pt}\discretionary{.}{%
}{.}\hspace{.4pt}2196\discretionary{/}{%
}{/}34781}}}


\bibitem{qinghong2022egocentric}
K.~Qinghong~Lin, A.~Jinpeng~Wang, M.~Soldan, M.~Wray, R.~Yan, E.~Zhongcong~Xu,
  D.~Gao, R.~Tu, W.~Zhao, W.~Kong, et~al.
\newblock Egocentric video-language pretraining.
\newblock {\em arXiv e-prints}, pp. arXiv--2206, 2022.

\bibitem{radford2021learning}
A.~Radford, J.~W. Kim, C.~Hallacy, A.~Ramesh, G.~Goh, S.~Agarwal, G.~Sastry,
  A.~Askell, P.~Mishkin, J.~Clark, et~al.
\newblock Learning transferable visual models from natural language
  supervision.
\newblock In {\em International conference on machine learning}, pp.
  8748--8763. PMLR, 2021.

\bibitem{fast_api}
S.~Ramírez.
\newblock Fastapi.
\newblock \url{https://fastapi.tiangolo.com/}.

\bibitem{schmeil_mara_2007}
A.~Schmeil and W.~Broll.
\newblock {MARA} - {A} {Mobile} {Augmented} {Reality}-{Based} {Virtual}
  {Assistant}.
\newblock In {\em 2007 {IEEE} {Virtual} {Reality} {Conference}}, pp. 267--270,
  2007. \href{https://doi.org/10.1109/VR.2007.352497}
{doi: {{%
10\hspace{.1pt}\discretionary{.}{%
}{.}\hspace{.4pt}1109\discretionary{/}{%
}{/}VR\hspace{.1pt}\discretionary{.}{%
}{.}\hspace{.4pt}2007\hspace{.1pt}\discretionary{.}{%
}{.}\hspace{.4pt}352497}}}


\bibitem{shneiderman1996eyes}
B.~Shneiderman.
\newblock The eyes have it: A task by data type taxonomy for information
  visualizations.
\newblock In {\em Proceedings 1996 IEEE symposium on visual languages}, pp.
  336--343. IEEE, 1996.

\bibitem{sicat_dxr_2019}
R.~Sicat, J.~Li, J.~Choi, M.~Cordeil, W.-K. Jeong, B.~Bach, and H.~Pfister.
\newblock {DXR}: {A} {Toolkit} for {Building} {Immersive} {Data}
  {Visualizations}.
\newblock {\em IEEE Transactions on Visualization and Computer Graphics},
  25(1):715--725, 2019.
\newblock Conference Name: IEEE Transactions on Visualization and Computer
  Graphics. \href{https://doi.org/10.1109/TVCG.2018.2865152}
{doi: {{%
10\hspace{.1pt}\discretionary{.}{%
}{.}\hspace{.4pt}1109\discretionary{/}{%
}{/}TVCG\hspace{.1pt}\discretionary{.}{%
}{.}\hspace{.4pt}2018\hspace{.1pt}\discretionary{.}{%
}{.}\hspace{.4pt}2865152}}}


\bibitem{react}
M.~O. Source.
\newblock React.
\newblock \url{https://react.dev/}.

\bibitem{1544870}
D.~Steedly, C.~Pal, and R.~Szeliski.
\newblock Efficiently registering video into panoramic mosaics.
\newblock In {\em Tenth IEEE International Conference on Computer Vision
  (ICCV'05) Volume 1}, vol.~2, pp. 1300--1307 Vol. 2, 2005.
  \href{https://doi.org/10.1109/ICCV.2005.86}
{doi: {{%
10\hspace{.1pt}\discretionary{.}{%
}{.}\hspace{.4pt}1109\discretionary{/}{%
}{/}ICCV\hspace{.1pt}\discretionary{.}{%
}{.}\hspace{.4pt}2005\hspace{.1pt}\discretionary{.}{%
}{.}\hspace{.4pt}86}}}


\bibitem{sun_high-precision_2020}
X.~Sun, S.~B. Murthi, G.~Schwartzbauer, and A.~Varshney.
\newblock High-{Precision} {5DoF} {Tracking} and {Visualization} of {Catheter}
  {Placement} in {EVD} of the {Brain} {Using} {AR}.
\newblock {\em ACM Transactions on Computing for Healthcare}, 1(2):9:1--9:18,
  2020. \href{https://doi.org/10.1145/3365678}
{doi: {{%
10\hspace{.1pt}\discretionary{.}{%
}{.}\hspace{.4pt}1145\discretionary{/}{%
}{/}3365678}}}


\bibitem{tang2003comparative}
A.~Tang, C.~Owen, F.~Biocca, and W.~Mou.
\newblock Comparative effectiveness of augmented reality in object assembly.
\newblock In {\em Proceedings of the SIGCHI conference on Human factors in
  computing systems}, pp. 73--80, 2003.

\bibitem{3js}
three.js.
\newblock three.js.
\newblock \url{https://threejs.org/}.

\bibitem{DBLP:journals/corr/abs-2008-11239}
D.~Ungureanu, F.~Bogo, S.~Galliani, P.~Sama, X.~Duan, C.~Meekhof,
  J.~St{\"{u}}hmer, T.~J. Cashman, B.~Tekin, J.~L. Sch{\"{o}}nberger,
  P.~Olszta, and M.~Pollefeys.
\newblock Hololens 2 research mode as a tool for computer vision research.
\newblock {\em CoRR}, abs/2008.11239, 2020.

\bibitem{xiao2020audiovisual}
F.~Xiao, Y.~J. Lee, K.~Grauman, J.~Malik, and C.~Feichtenhofer.
\newblock Audiovisual slowfast networks for video recognition.
\newblock {\em arXiv preprint arXiv:2001.08740}, 2020.

\bibitem{zhang_manifold_2019}
J.~Zhang, Y.~Wang, P.~Molino, L.~Li, and D.~S. Ebert.
\newblock Manifold: {A} {Model}-{Agnostic} {Framework} for {Interpretation} and
  {Diagnosis} of {Machine} {Learning} {Models}.
\newblock {\em IEEE Transactions on Visualization and Computer Graphics},
  25(1):364--373, 2019. \href{https://doi.org/10.1109/TVCG.2018.2864499}
{doi: {{%
10\hspace{.1pt}\discretionary{.}{%
}{.}\hspace{.4pt}1109\discretionary{/}{%
}{/}TVCG\hspace{.1pt}\discretionary{.}{%
}{.}\hspace{.4pt}2018\hspace{.1pt}\discretionary{.}{%
}{.}\hspace{.4pt}2864499}}}


\bibitem{zhang2012automatically}
Z.~Zhang, P.~Webster, V.~S. Uren, A.~Varga, and F.~Ciravegna.
\newblock Automatically extracting procedural knowledge from instructional
  texts using natural language processing.
\newblock In {\em LREC}, vol. 2012, pp. 520--527. Citeseer, 2012.

\bibitem{zheng2015eye}
X.~S. Zheng, C.~Foucault, P.~Matos~da Silva, S.~Dasari, T.~Yang, and S.~Goose.
\newblock Eye-wearable technology for machine maintenance: Effects of display
  position and hands-free operation.
\newblock In {\em Proceedings of the 33rd Annual ACM Conference on Human
  Factors in Computing Systems}, pp. 2125--2134, 2015.

\bibitem{zhou2021domain}
K.~Zhou, Z.~Liu, Y.~Qiao, T.~Xiang, and C.~C. Loy.
\newblock Domain generalization in vision: A survey.
\newblock {\em arXiv preprint arXiv:2103.02503}, 2021.

\bibitem{zhou2022detecting}
X.~Zhou, R.~Girdhar, A.~Joulin, P.~Kr{\"a}henb{\"u}hl, and I.~Misra.
\newblock Detecting twenty-thousand classes using image-level supervision.
\newblock In {\em Computer Vision--ECCV 2022: 17th European Conference, Tel
  Aviv, Israel, October 23--27, 2022, Proceedings, Part IX}, pp. 350--368.
  Springer, 2022.

\end{thebibliography}

% \newpage
% \appendix
% \input{text/08-appendix}

%% ^^^^^   FOR IEEE VIS, EVERYTHING HERE MAY BE INCLUDED IN THE    ^^^^^ %%
%% 2-PAGE ALLOTMENT FOR REFERENCES, FIGURE CREDITS, AND ACKNOWLEDGEMENTS %%

\end{document}